\documentclass[%
 reprint,
superscriptaddress,
 amsmath,amssymb,
 aps,
]{revtex4-2}
\usepackage{comment}
\usepackage{graphicx}
\usepackage{dcolumn}
\usepackage{bm}

\usepackage{xcolor}
\usepackage{units}
\usepackage{appendix}
\usepackage{float}
\usepackage{dsfont}
\usepackage{soul}

\usepackage[version=4]{mhchem}
\usepackage{tabularx}
\usepackage{multirow}
\usepackage{soul}
\usepackage{subfig}
\usepackage[colorlinks = true, linkcolor = blue, urlcolor  = blue, citecolor = blue, anchorcolor = blue]{hyperref}

\begin{document}
\preprint{APS/123-QED}

\title{Spin-lattice couplings in $3d$ ferromagnets: analysis from first-principles}

\author{I. P. Miranda}
\affiliation{Department of Physics and Astronomy, Uppsala University, Box 516, SE-75120 Uppsala, Sweden}
\affiliation{Department of Physics and Electrical Engineering, Linnaeus University, SE-39231 Kalmar, Sweden}
\email[Corresponding author:\ ]{ivan.miranda@alumni.usp.br}

\author{M. Pankratova}
\affiliation{Department of Engineering Sciences, University of Skövde, SE-541 28 Skövde, Sweden}
\affiliation{Department of Physics and Astronomy, Uppsala University, Box 516, SE-75120 Uppsala, Sweden}
\affiliation{Wallenberg Initiative Materials Science for Sustainability, Department of Physics and Astronomy, Uppsala University, 751 21 Uppsala, Sweden}

\author{M. Weißenhofer}
\affiliation{Department of Physics and Astronomy, Uppsala University, Box 516, SE-75120 Uppsala, Sweden}
\affiliation{Department of Physics, Freie Universit{\"a}t Berlin, Arnimallee 14, D-14195 Berlin, Germany}

\author{A. B. Klautau}
\affiliation{Faculdade de F\'isica, Universidade Federal do Par\'a, CEP 66075-110, Bel\'em, PA, Brazil}

\author{D. Thonig}
\affiliation{School of Science and Technology,  \"Orebro University, SE-701 82, \"Orebro, Sweden}
\affiliation{Department of Physics and Astronomy, Uppsala University, Box 516, SE-75120 Uppsala, Sweden}

\author{M. Pereiro}
\affiliation{Department of Physics and Astronomy, Uppsala University, Box 516, SE-75120 Uppsala, Sweden}

\author{E. Sjöqvist}
\affiliation{Department of Physics and Astronomy, Uppsala University, Box 516, SE-75120 Uppsala, Sweden}

\author{A. Delin}
\affiliation{Department of Applied Physics, School of Engineering Sciences, KTH Royal Institute of Technology, AlbaNova University Center, SE-10691 Stockholm, Sweden}
\affiliation{ Swedish e-Science Research Center (SeRC), KTH Royal Institute of Technology, SE-10044 Stockholm, Sweden}
\affiliation{Wallenberg Initiative Materials Science for Sustainability (WISE), KTH Royal Institute of Technology, SE-10044 Stockholm, Sweden}

\author{M. I. Katsnelson}
\affiliation{Institute for Molecules and Materials, Radboud University, Heijendaalseweg 135, 6525 AJ Nijmegen, The Netherlands}
\affiliation{Wallenberg Initiative Materials Science for Sustainability, Department of Physics and Astronomy, Uppsala University, 751 21 Uppsala, Sweden}

\author{O. Eriksson}
\affiliation{Department of Physics and Astronomy, Uppsala University, Box 516, SE-75120 Uppsala, Sweden}
\affiliation{Wallenberg Initiative Materials Science for Sustainability, Department of Physics and Astronomy, Uppsala University, 751 21 Uppsala, Sweden}

\author{A. Bergman}
\affiliation{Department of Physics and Astronomy, Uppsala University, Box 516, SE-75120 Uppsala, Sweden}

\date{\today}

\begin{abstract}
Magnetoelasticity plays a crucial role in numerous magnetic phenomena, including magnetocalorics, magnon excitation via acoustic waves, and ultrafast demagnetization/Einstein-de Haas effect. Despite a long-standing discussion on anisotropy-mediated magnetoelastic interactions of relativistic origin, the exchange-mediated magnetoelastic parameters within an atomistic framework have only recently begun to be investigated. As a result, many of their behaviors and values for real materials remain poorly understood. Therefore, by using a proposed simple modification of the embedded cluster approach that reduces the computational complexity, we critically analyze the properties of exchange-mediated spin-lattice coupling parameters for elemental $3d$ ferromagnets (bcc Fe, fcc Ni, and fcc Co), comparing methods used for their extraction and relating their realistic values to symmetry considerations and orbitally-decomposed contributions. Additionally, we investigate the effects of noncollinearity (spin temperature) and applied pressure on these parameters. For Fe, we find that single-site rotations, associated with spin temperatures around $\sim100$ K, induce significant modifications, particularly in Dzyaloshinskii-Moriya-type couplings; in contrast, such interactions in Co and Ni remain almost configuration independent. Moreover, we demonstrate a notable change in the exchange-mediated magnetoelastic constants for Fe under isotropic contraction. Finally, the conversion between atomistic, quantum-mechanically derived parameters and the phenomenological magnetoelastic theory is discussed, which can be an useful tool towards larger and more realistic dynamics simulations involving coupled subsystems.

\end{abstract}

\maketitle

\section{INTRODUCTION} 
Spin-lattice is a fundamental interaction, which, despite its first theoretical description dating from about seven decades ago \cite{Kittel1949,Kaganov1959,Abrahams1952}, is attracting ever-increasing attention. Its importance has been demonstrated in multiple phenomena, including ultrafast demagnetization \cite{beaurepaire1996ultrafast}, magnetocaloric effect \cite{Patra2023}, magnon excitation via acoustic waves \cite{Li2021}, thermal Hall effect \cite{Zhang2019}, and skyrmion transport \cite{Wu2021} -- to cite a few. In the context of ultrafast demagnetization, which serves as a prominent example, it was shown that during pump-probe experiments the lost spin angular momentum is constantly transferred (via spin-lattice coupling) to the lattice subsystem, originating circularly polarized phonons that later evolve to the macroscopic Einstein-de Haas effect \cite{Lemeshko,tauchert2022polarized,Dornes2019}. It was also demonstrated in the pioneering work of Beaurepaire \cite{beaurepaire1996ultrafast} (and derived temperature models \cite{Pankratova2022, Pankratova2023}) that the values of the coefficient $G_{sl}$, responsible for the transfer of heat between spin and lattice subsystems, can influence the occurrence of a maximum in the spin temperature, and therefore, a minimum in the demagnetization curve. 

Another examples, which demonstrate the relevance of spin-lattice coupling from the perspective of spintronics, are (\textit{i}) the impact of spin-lattice coupling on skyrmions dynamics \cite{wu2021microdynamic}, and (\textit{ii}) the generation of a thermal Hall effect \cite{Zhang2019}.
In addition, the importance of accounting of spin-lattice coupling for two-dimensional systems was demonstrated in \cite{PhysRevB.105.104418}, where it was shown that the magnetic interactions in CrI$_3$ are sensitive to atomic displacements; in fact, for large enough displacements, the exchange interactions can be changed from ferromagnetic to antiferromagnetic. Phonon contribution to exchange parameters in the other two-dimensional magnet, Fe$_3$GeTe$_2$ was also shown to be essential leading to noticeable renormalization of magnon spectrum and reduction of Curie temperature by $\approx 10\%$ \cite{Rudenko}.  
Recently, it was also demonstrated in \cite{zhu2017modulation} that voltage control of magnetism can be achieved in multiferroic heterostructures. The results were interpreted using a spin-lattice coupling model. Specifically, it was shown that variation of exchange stiffness is connected to voltage-induced anisotropic lattice changes. Using spin-lattice coupling to control spin dynamics can be useful from the perspective of energy-efficient, tunable magnonics devices. One more possibility to exploit spin-lattice to control magnetisation was discussed in \cite{dos2023spin}, where using coupled spin-lattice dynamics simulations, the effects of pressure on the magnetic properties were investigated. It was shown that the effect of pressure is more pronounced near the Curie temperature, and that magnetization can either increase or decrease depending on the type of contraction. 

The importance of accounting for spin-lattice coupling for the mentioned and many other phenomena \cite{Lange2023} led to growing efforts to its estimation or calculation via first-principles. As discussed above, Beaurepaire \textit{et al.} \cite{beaurepaire1996ultrafast} estimated the spin-lattice coupling parameter in Ni by fitting the experimental data with a three-temperature model. Later, spin-lattice started to be indirectly calculated via density functional theory; since the pioneering works of Ma \textit{et al.} \cite{Ma2008,Ma2012}, such interactions begin to be considered in an atomistic perspective, moving beyond the macroscopic quantities traditionally described by the classical magnetoelastic theory \cite{Kittel1949}. In Refs. \cite{Ma2008,Ma2012} (and many others that later incorporated similar coupled dynamics), the spin-lattice interactions were simply treated by a fitting of the exchange couplings as a function of the scalar distance, $J_{ij}(r_{ij})$ -- considering the $J_{ij}$ values obtained with the lattice at equilibrium positions. More recently, and still within the scope of atomistic modeling, Ref. \cite{Hellsvik2019} developed a method of calculating spin-lattice via a first-order derivative of the spin tensor with respect to atomic displacements. This was a relevant step, since it allowed for accurate (in terms of predictions via electronic structure) and spatially-dependent results, with the disadvantage of higher computational cost. Aiming for a solution to that problem, and in view of the application of similar calculations as demonstrated in \cite{Hellsvik2019} to more complex systems, a few alternative approaches were later developed. Among the proposed methods we can cite: the supercell approach \cite{Mankovsky2022,Lange2023}, the embedded cluster approach \cite{Lange2023}, and the perturbative approach \cite{Mankovsky2022,Lange2023}. Lastly, some machine learning-based methods to include spin-lattice coupling were also recently proposed for large-scale spin-lattice dynamics simulations, as reported in  \cite{nikolov2021data} (and references therein). 

Despite such initial efforts to calculate spin-lattice coupling, progress in this field remains in its early stages \cite{Lange2023}. Developing a reliable and accurate method for calculating spin-lattice coupling that is applicable across a wide range of materials is highly desirable for advancing theoretical materials modeling. At the same time, while the magnetoelastic terms introduced early by Kittel and Abrahams \cite{Kittel1949,Abrahams1952} and Kaganov and Tsukernik \cite{Kaganov1959} have been extensively studied in the literature, the properties of microscopic (atomistic), quantum-mechanically derived, spin-lattice interactions -- and their connection to measurable quantities within the classical magnetoelastic theory -- remain poorly understood. Here, we aim to fill this gap by providing a comprehensive investigation of those listed points, as well as the behavior of the spin-lattice interactions under external factors such as temperature and applied pressure, with \textit{ab-initio} accuracy.

The structure of the paper is as follows. First, in Section \ref{sec:theory}, we revisit existing spin-lattice coupling calculation methods and propose a simple (but crucial) modification that reduces the computational complexity, allowing for faster and reliable  calculations. Then, in Section \ref{sec:analysis-3d-ferromagnets}, we calculate, compare the spin-lattice coupling parameters for $3d$ ferromagnets using various approaches, and analyze their properties. Later, in Section \ref{sec:temperature}, we estimate the impact of the noncollinear arrangement of spins due to, for example, thermal effects on the spin-lattice interactions. Finally, in Section \ref{sec:pressure-effects}, we investigate the pressure effects for the case of bcc Fe, aiming for an estimation of the tunability of the spin-lattice interactions via an external stimulus. 

\section{THEORY}
\label{sec:theory}

\subsection{Spin Hamiltonian}

As suggested by and reviewed in many previous works, the Heisenberg exchange, Dzyaloshinskii-Moriya interactions and symmetric anisotropic exchange are all belonging to the so-called bilinear spin couplings. Thus, they can be written in the form of a compact, generalized exchange tensor $\mathcal{J}_{ij}$ to which the corresponding Hamiltonian term is expressed as \cite{Hellsvik2019} (with Einstein notation applied for double occurring Greek indices)

\begin{equation}
\label{eq:generalized-bilinar-hamiltonian}
\mathcal{H}_{SS}=-\frac{1}{2}\sum_{ij}\mathcal{J}_{ij}^{\alpha\beta}m_i^{\alpha}m_j^{\beta},
\end{equation}

\noindent where $\alpha,\beta=\{x,y,z\}$ are Cartesian indices, $m_i^{\alpha}$ is the local spin magnetic moment direction $\alpha$ of the $i$-th atom (represented here as a unit vector), and $i$ and $j$ are site indices. The $\frac{1}{2}$ factor is introduced to avoid double counting in the summation. It is well-known that the quantities in the $\mathcal{J}_{ij}$ tensor have a spatial dependence. This means that atomic displacements $u_k^{\mu}$ in the lattice of a $k$-th atom (with $\mu=\{x,y,z\}$) will alter those quantities as $\mathcal{J}_{ij}(u_k^{\mu})$, by changing the electronic structure as a whole. Fortunately, in real materials, the displacements $u_k^{\mu}$ due to thermal effects in the lattice are usually small, agreeing with the famous Lindemann criterion \cite{Lindemann1910} \footnote{The Lindemann criterion states that the melting of a three-dimensional solid occurs when the square root of the mean-squared displacement reaches a threshold of $\sim10\%$ of the interparticle distance, with respect to the equilibrium positions.}, recent theoretical works \cite{Pankratova2022}, and experiments. This suggests that the tensor elements $\mathcal{J}_{ij}^{\alpha\beta}(u_k^{\mu})$ can be expanded in a Taylor series (in which just the first term is assumed to be relevant) \cite{Hellsvik2019}

\begin{equation}
\label{eq:taylor-expansion}
\mathcal{J}_{ij}^{\alpha\beta}(u_k^{\mu})\approx\mathcal{J}_{ij}^{\alpha\beta}(0)+\frac{\partial\mathcal{J}_{ij}^{\alpha\beta}}{\partial u_k^{\mu}}u_k^{\mu},
\end{equation}

\noindent where $\mathcal{J}_{ij}(0)$ is the tensor calculated for lattice in the equilibrium positions. The resulting rank-3 tensor with elements $\Gamma_{ijk}^{\alpha\beta\mu}=\frac{\partial\mathcal{J}_{ij}^{\alpha\beta}}{\partial u_k^{\mu}}$ is what we refer to as spin-lattice, composed by the direct product of a rank-2 tensor in spin space and a rank-1 tensor in orbital space. Thus, the spin-spin Hamiltonian is rewritten as 

\begin{equation}
\label{eq:spin-lattice-hamiltonian-at}
\mathcal{H}_{SS}\approx -\frac{1}{2}\sum_{ij}\mathcal{J}_{ij}^{\alpha\beta}(0)m_i^{\alpha}m_j^{\beta} -\frac{1}{2}\sum_{ijk}\frac{\partial\mathcal{J}_{ij}^{\alpha\beta}}{\partial u_k^{\mu}}m_i^{\alpha}m_j^{\beta}u_k^{\mu}.
\end{equation}

The diagonal parts of both the $\mathcal{J}_{ij}(0)$ and the $\frac{\partial\mathcal{J}_{ij}
}{\partial u_k^{\mu}}$ tensors, namely when $\alpha=\beta$, are referred to as the Heisenberg exchange (at the equilibrium positions) and the Heisenberg-like spin-lattice parameters, respectively. For those particular values, the symbol $J_{ij}$ will be used instead, so that the definition $J_{ij}=\frac{1}{2}\left(\mathcal{J}_{ij}^{xx}+\mathcal{J}_{ij}^{yy}\right)$ (for $\hat{z}$ as the spin quantization axis) holds \cite{Udvardi2003}. Particularly in crystals with cubic symmetry, we have simply $J_{ij}=\mathcal{J}_{ij}^{\alpha\alpha}$ and $\Gamma_{ijk}^{\alpha\alpha\mu}=\frac{\partial\mathcal{J}_{ij}^{\alpha\alpha}}{\partial u_k^{\mu}}=\frac{\partial J_{ij}}{\partial u_k^{\mu}}$. This is not the case for more complex materials, such as CrI$_3$ \cite{Kvashnin2020,Sadhukhan2022}. 

\subsection{Determination of bilinear spin-lattice couplings}

In this section, we revise two established approaches from the literature to calculate the $\frac{\partial\mathcal{J}_{ij}^{\alpha\beta}}{\partial u_k^{\mu}}$ parameters, and propose a simplification of the embedded cluster approach within the linear muffin-tin orbital in the atomic sphere approximation (LMTO-ASA) method \cite{Andersen1975}. We note that these are not the only methods available; another approach is, for instance, the so-called \textit{perturbative}, recently described in Ref. \cite{Mankovsky2022}.

\subsubsection{Supercell and embedded cluster approaches}
\label{sec:eca}
The immediate \textit{ab-initio} method to calculate the interaction between spin and lattice degrees of freedom is the self-consistent determination of $J_{ij}(\mathbf{u}_k)$, by explicitly displacing the $k$-th atom by a certain amount $\left|\mathbf{u}_{k}\right|$, proceeding then to the determination of the new self-consistent (perturbed) charge density. As the most straightforward approach, this was suggested by Hellsvik \textit{et al.} \cite{Hellsvik2019}, and employed to obtain the spin-lattice coupling parameters of bcc Fe. In order to correctly describe the local perturbation on the charge density, and avoid the presence of spurious interactions from periodic images, it is common to construct a supercell around the displaced atom $k$ \cite{Mankovsky2022, Lange2023}. As noticed by Lange \textit{et al.} \cite{Lange2023}, the self-consistent nature of this approach comes with the advantage of accuracy (for a sufficient supercell size) and disadvantage of generally high computational cost.

In real space, however, spurious interactions are in principle absent by construction, so that the perturbation in the charge density can be well-described by considering a big enough cluster around the atom $k$, which will also capture the charge transfer with the neighboring sites -- the embedded cluster approach (ECA). To address this, one can conceptualize the cluster as an impurity problem \cite{Frota-Pessoa1992}. Initially, two methods can be employed: (\textit{i}) solve for the atoms within the cluster self-consistently, wherein all external atoms maintain the potential parameters of the perfect crystal; or (\textit{ii}) resolve using the Dyson equation \cite{Lange2023}, considering this cluster as a local interaction zone \cite{Abrikosov1997}. In particular, for small displacements $\left|\mathbf{u}_{k}\right|$, and given the strong dependence of the LMTO Hamiltonian with the system structure (through the so-called structure matrix $\mathbf{S}$ \cite{Andersen1984}), one can outline the following approximation for the fully self-consistent ECA:

\begin{itemize}
    \item after displacing the desired atom from the equilibrium position and choosing a suitable cluster around it, we recalculate the electrostatic potential and the matrix $\mathbf{S}$ \cite{Frota-Pessoa1992,Andersen1984}, keeping the potential parameters unchanged as in the perfect crystal situation;
    \item we then proceed to a single iteration, and by the determination of $J_{ij}(\mathbf{u}_k)$ using the well-known Liechtenstein-Katsnelson-Antropov-Gubanov (LKAG) \cite{Liechtenstein1987,Szilva2023} formalism;
    \item for the subsequent displacement, we start from \textit{this} (already displaced) result and repeat the previous steps, provided that the difference $\Delta \mathbf{u}_k$ to the next calculation is quite small.
\end{itemize}

This simplification (which we will refer to as \textit{sECA}) allows for a significant reduction of the computational effort involved in the ECA with the cost of lower accuracy, as the perturbed charge density is not permitted to fully relax. However, as it will be discussed in the Section \ref{sec:results}, the fact that displacements are small allows for a very good agreement between those two methods for the investigated cases here. Due to its inherent simplicity, the sECA enables the calculation of spin-lattice parameters in complex structures.

\subsubsection{Determination by fitting}
\label{sec:determination-by-fitting}

Recent works on the spin-lattice dynamics are based on a scalar dependence of the exchange interactions with inter-atomic distances $r_{ij}$ in their equilibrium positions, $J_{ij}({r}_{ij})$ \cite{Assmann2019,Strungaru2021,Nieves2021,Zhou2020,Nikolov2021,Patra2023,Korniienko2024}, specially since the investigations by Ma \textit{et al.} \cite{Ma2008,Ma2012}. The commonly used fitting functions are assumed to have a shape based either on the Bethe-Slater curve \cite{Yosida1996} or present an oscillatory behavior (which reproduces the Ruderman-Kittel-Kasuya-Yosida  \cite{Scheie2022} (RKKY)-like nature), with an exponential decay \cite{Assmann2019}.  However, despite the obvious computational efficiency, these approximations account for a long-range nature of the exchange function in a crystal lattice (relatively to the length of realistic displacements $u_k^{\mu}$) to calculate the change in $J_{ij}$ due to small perturbations $u_k^{\mu}$, assuming that all neighbors interactions will vary in the same way.  

In order to make a comparison with other spin-lattice results in literature, we define three functions ($f_1$, $f_2$, $f_3$) based on $r_{ij}$ that are commonly used \cite{Assmann2019,Strungaru2021,Nieves2021,Zhou2020,Nikolov2021,Ma2008,Ma2012,Patra2023}, (see Section \ref{sec:analysis-3d-ferromagnets}):

\begin{equation}\label{eq:modelsjij}
\begin{split}
f_{1}(r_{ij})=J_{0}\left[\frac{C_{1}\cos(\omega(r_{ij}-d_1))}{1+\exp(\xi(r_{ij}-d_1)}+C_{2}\exp(-\gamma(r_{ij}-d_2)^{2})\right], \\
f_{2}(r_{ij})=J_0\left(1-\frac{r_{ij}}{r_{c}}\right)^{3}\Theta(r_{c}-r_{ij}),\\
f_{3}(r_{ij})=4\alpha_{J}\left(\frac{r_{ij}}{\delta_J}\right)^{2}\left[1-\gamma_{n}\left(\frac{r_{ij}}{\delta_J}\right)^{2}\right]e^{-\left(\frac{r_{ij}}{\delta_J}\right)^{2}}\Theta(r_{c}-r_{ij}),
\end{split}
\end{equation}

\noindent where $r_c$ is defined as a cutoff radius, $\Theta$ is the Heaviside step function, $J_{0}$ a Heisenberg-like exchange constant, and all other parameters to be fitted with \textit{ab-initio} values.

Another formulation of the fitting procedure, which also uses the functions defined in Eq. \ref{eq:modelsjij}, is based on an effective short-range exchange interaction \cite{Nieves2021}. In this approach, the exchange interactions are considered up to nearest neighbors, and the direct relation to volume magnetostriction and elastic constants is used to calculate $\frac{\partial J}{\partial r}$. In this model, the experimental magnetostriction can be well-reproduced by construction. However, clear limitations include a higher dependence of input parameters, and the omission of longer-range  spin-lattice interactions, which has shown to be important in, e.g., bcc Fe \cite{Wang2010}.

\section{COMPUTATIONAL DETAILS}
\label{sec:computational}

The parameterization of Eq. \ref{eq:spin-lattice-hamiltonian-at} in terms of calculation of $\mathcal{J}_{ij}^{\alpha\beta}$ and realistic $u_k$ displacements was obtained in the framework of the Density Functional Theory (DFT), combining two different methods: (\textit{i}) the real-space linear muffin-tin orbital in the atomic sphere approximation (RS-LMTO-ASA) \cite{Frota-Pessoa1992}, and (\textit{ii}) the plane-wave pseudopotential-based Quantum ESPRESSO (QE) package \cite{Giannozzi2017}, in combination with PHONOPY \cite{Togo2015}. The hybrid choice of methods is justified by the more suitable way of treating the spin-lattice coupling (SLC) parameters in a broken inversion symmetry environment by RS-LMTO-ASA when an atom is displaced. However, the Hellman-Feynman forces are known to be inaccurately described by this method \cite{Wills2010}. To overcome this problem, the realistic $u_k$ values are calculated from the force constants, using a combination between QE and PHONOPY. 

In the real-space formalism of RS-LMTO-ASA, the eigenvalue problem is solved with the help of the recursion method, where the recursion chain is ended after $LL$ steps by making use of the Beer-Pettifor terminator \cite{Haydock1980,Beer1984}. For the calculation of the electronic structure of all $3d$ ferromagnets investigated here (Fe, Co and Ni), we considered $LL=31$. The linear approach provides precise results around an energy $E_{\nu}$, which is typically chosen as the center of gravity of the occupied bands. Here, we work in the orthogonal representation of the LMTO-ASA expanded in terms of tight-binding parameters \cite{Andersen1984,Andersen1975}, neglecting terms of the order $(E-E_{\nu})^3$ and higher. The structures were simulated by big clusters whose atoms are located in the perfect bcc/fcc crystal positions, with experimental lattice parameters of $a_{\textnormal{Fe}}=2.87\,$\AA$\,$, $a_{\textnormal{Co}}=3.54\,$\AA$\,$, and $a_{\textnormal{Ni}}=3.52\,$\AA$\,$ \cite{Lu2023}. In this setup, the local density approximation (LSDA), with the von Barth and Hedin parametrization \cite{Barth1972}, was used. When necessary, the spin-orbit coupling is included as a $\mathbf{L}\cdot\mathbf{S}$ term to be computed at each variational step \cite{Andersen1975}. 

Still in the context of (\textit{i}), in both embedded cluster methods, the exchange parameters $\mathcal{J}^{\alpha\alpha}_{ij}(\textbf{u}_k)\equiv J_{ij}(\textbf{u}_k)$ were obtained by the LKAG approach \cite{Liechtenstein1987,Frota-Pessoa2000} for each finite displacement $\textbf{u}_k$:

\begin{equation}
\label{eq:lkag}
J_{ij} = \frac{1}{4\pi}\textnormal{ImTr}\int_{-\infty}^{E_F}\boldsymbol{\delta}_{i}\mathbf{G}_{ij}^{\uparrow\uparrow}\boldsymbol{\delta}_{j}\mathbf{G}_{ji}^{\downarrow\downarrow}d\epsilon,
\end{equation}

\noindent where $\boldsymbol{\delta}_i$ is an exchange-splitting matrix and $\mathbf{G}_{ij}^{\varsigma\varsigma^{\prime}}$ are intersite Green's functions between the $\varsigma$-$\varsigma^{\prime}$ spin channels. The calculation of Dzyaloshinskii-Moriya vectors follow analogous expressions, whose derivations are referred to Ref. \cite{Cardias2020,Udvardi2003}.

As suggested by Ref. \cite{Fransson2017}, the single-particle Green's functions can be rewritten in terms of charge (0) and spin (1) components as $\mathbf{G}=(G^{00}+G^{01})\sigma_{0}+(\mathbf{G}^{10}+\mathbf{G}^{11})\cdot\boldsymbol{\sigma}$, with $\boldsymbol{\sigma}=(\sigma_x, \sigma_y, \sigma_z)$ the vector of Pauli matrices, and $\sigma_0$ the $2\times2$ identity matrix. Thus, when in a (more general) noncollinear situation, the exchange interactions can be determined as

\begin{equation}
\label{eq:lkag-general}
\begin{split}
J_{ij}=\frac{1}{4\pi}\textnormal{ImTr}\int_{-\infty}^{E_F}(\boldsymbol{\delta}_{i}{G}_{ij}^{00}\boldsymbol{\delta}_{j}{G}_{ji}^{00}+\boldsymbol{\delta}_{i}{G}_{ij}^{01}\boldsymbol{\delta}_{j}{G}_{ji}^{01}-\\\boldsymbol{\delta}_{i}\mathbf{G}_{ij}^{10}\cdot\boldsymbol{\delta}_{j}\mathbf{G}_{ji}^{10}-\boldsymbol{\delta}_{i}\mathbf{G}_{ij}^{11}\cdot\boldsymbol{\delta}_{j}\mathbf{G}_{ji}^{11})d\epsilon,
\end{split}
\end{equation}

\noindent with analogous expressions for the off-diagonal terms of the generalized exchange tensor (Eq. \ref{eq:generalized-bilinar-hamiltonian}) \cite{Fransson2017}.

The realistic magnitude of atomic displacements $\mathbf{u}_k$ (as well as the equation of state discussed in Appendix \ref{sec:appendix-pressure}) were computed using a combination between QE and PHONOPY. For that, a scalar relativistic projected augmented wave (PAW) pseudopotential was used. In all $3d$ ferromagnets, the plane waves were expanded for a kinetic energy and charge density cutoffs of 60-100 Ry and 480-1080 Ry, respectively. For the ground-state properties (primitive cell), the reciprocal space was sampled for unshifted $\mathbf{k}$-meshs ranging from  $18\times18\times18$ to $24\times24\times24$, depending on the material (specific details in Refs. \cite{Pankratova2022,Pankratova2023}). All PHONOPY calculations were based in a $6\times6\times6$ supercell (216 atoms), for which reduced $\mathbf{k}$-meshs were used.

Note that the thermally average effects of spin-lattice coupling to exchange parameters can be calculated in a much simpler and more straightforward way \cite{Rudenko}, by adding electron-phonon contribution to the electron self energy in equations like (\ref{eq:lkag-general}). However, to study spin and lattice dynamics and out-of-equilibrium phenomena in general much more detailed information on the whole $\Gamma$-tensor is needed. 

\section{RESULTS}
\label{sec:results}

\subsection{Analysis of spin-lattice coupling in $3d$ ferromagnets}
\label{sec:analysis-3d-ferromagnets}

\subsubsection{Heisenberg-like parameters}

\begin{figure}[!htb]
\includegraphics[width=\columnwidth]{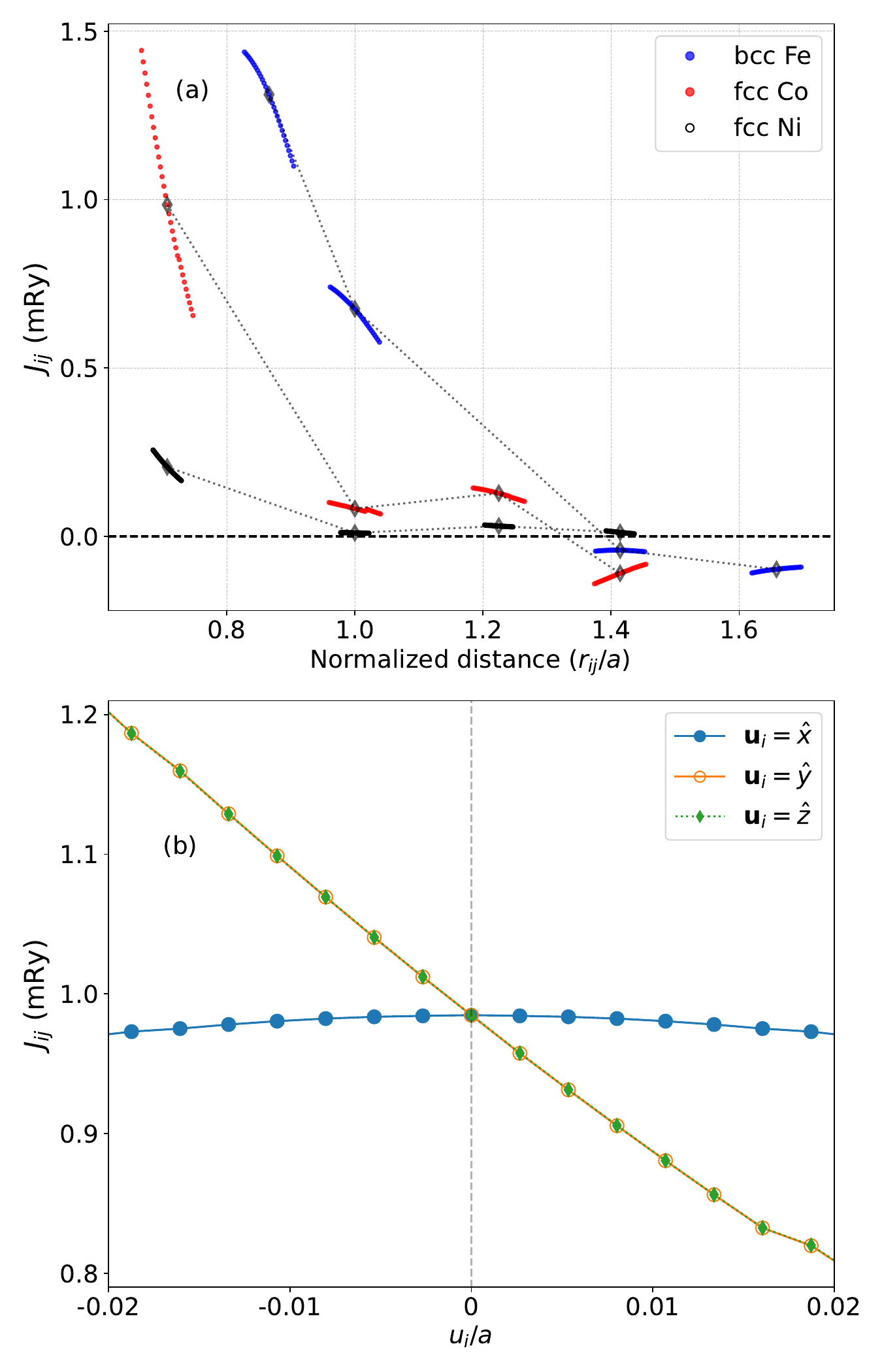}
\caption{(a) Variation of the Heisenberg exchange $J_{ij}$ for Fe, Co and Ni as a function of the $r_{ij}$ distance (in units of the lattice parameter) and displacement of the $i$-th atom. Black open diamonds represent the values for atoms in the equilibrium positions ($J_{ij}(0)$); (b) Variation of $J_{ij}$ for the case of nearest-neighbors in fcc Co, where $\mathbf{r}_{ij}=\left(0,\frac{1}{2},\frac{1}{2}\right)$, for distinct displacement directions $\mathbf{u}_i$. Dotted lines in (a) and solid lines in (b) are guides for the eyes.}
\label{fig:mml-heis}
\end{figure}

Although some discussion has been made in Refs. \cite{Pankratova2022,Pankratova2023}, in this section we revise the properties of $\frac{\partial\mathcal{J}_{ij}^{\alpha\beta}}{\partial u_k^{\mu}}$ for bcc Fe, fcc Co and fcc Ni, some of the simplest metallic ferromagnetic systems. 

Since the seminal paper in Ref. \cite{Hellsvik2019}, and subsequent investigations reported in Refs. \cite{Mankovsky2022,Lange2023,Mankovsky2023}, there seems to be a general agreement that $k=i$ (i.e., the displaced atom belongs to the $ij$ pair of the magnetic interaction) implies into the strongest diagonal SLC parameter (isotropic) contributions. This is understandable because the electronic intersite Green's function $\mathbf{G}_{ij}^{\varsigma}$ ($\varsigma=\uparrow,\downarrow$) will be directly affected by the reduced (or increased) distance $\mathbf{r}_{ij}^{\prime}=\mathbf{r}_{ij}\pm\mathbf{u}_{k=i}$ in Eq. \ref{eq:lkag}. In that sense, we start by analyzing the diagonal part of $\Gamma_{ijk}^{\alpha\beta\mu}$ for $k=i$, or $\Gamma_{iji}^{\alpha\alpha\mu}$, shown in Figure \ref{fig:mml-heis}(a)-(b). In all cases displayed, the spin quantization axis was set to $\hat{z}$ ($[001]$), and the exchange parameters were calculated for realistic displacements (Appendix \ref{sec:realistic-atomic-displacements}).

In Fig. \ref{fig:mml-heis}(a), a comparison between actual $J_{ij}$ values with respect to the atomic displacements for all three ferromagnets is shown. Here, two aspects can be highlighted: (\textit{i}) even if we follow the same displacement direction and convention, not always the exchange parameters decrease with increasing intersite distances (a strong example is the fourth neighborhood of fcc Co); and (\textit{ii}) for large $\mathbf{u}_i$ amplitudes, the $J_{ij}$ variation deviates from the linear behavior. While observation (\textit{ii}) just sets a natural limit to the $\mathbf{u}_i$ interval in which the first expansion term (Eq. \ref{eq:taylor-expansion}) dominates, that approximately coincides with the realistic values, observation (\textit{i}) illustrates an intrinsic connection of the spin-lattice parameters with the crystal structure. This is more clearly seen in Fig. \ref{fig:mml-heis}(b), where a distinct $J_{ij}$ variation is obtained depending on the direction of $\mathbf{u}_i$ in the nearest-neighborhood of fcc Co.

It is interesting to compare the trends in Fig. \ref{fig:mml-heis}(a) with those obtained in Ref. \cite{Wang2010}. In their article, Wang \textit{et al.} \cite{Wang2010} considered an isotropic contraction/expansion as a model for the spin-lattice coupling in bcc Fe. This approach generates a collective conduction-electron shift in the band positions and changes the Fermi surface topology, so that it induces a distinct behavior of first- and second-nearest-neighbors exchange interactions as a function of the distance (or the lattice parameter) as the one seen in Fig. \ref{fig:mml-heis}(a). In particular, an increase of $J_{ij}(r_{ij})$ is obtained near the equilibrium positions for both nearest- and second-nearest neighbors, implying in a positive volume magnetostriction -- as observed experimentally for iron \cite{Shimizu1978}.  In contrast, the model adopted here deals with local partial derivatives, measuring $\frac{\partial J_{ij}}{\partial\mathbf{u}_k}$ in a \textit{local sense}, while the rest of the environment is less perturbed.

Both observations are correct in their respective frameworks, yet differ in terms of scope and physical interpretation.
As discussed in Ref. \cite{Chikazumi2009}, in the context of magnetostriction alone, the main magnetoelastic effect of exchange interactions is the volume magnetostriction, to which Wang's approach is more appropriate to simulate. However, as we will explore later on, the atomic displacement method outlined here and in previous works (e.g., in \cite{Sabiryanov1999}) is more general: it provides both a volumetric magnetostriction effect and the local building blocks for the calculation of the magnetoelastic constants of the spin-gradient terms. The latter are  existent in more general magnetoelastic expressions, such as the one discussed by Kaganov and Tsukernik (KT) \cite{Kaganov1959}, and are essential for a comprehensive description of the interaction between spin waves and phonons.

As noted by previous works (see, e.g., Ref. \cite{Mankovsky2023}), contrary to Heisenberg exchange at the equilibrium positions, the elements of $\Gamma_{iji}^{\alpha\alpha\mu}$ are not only depending on the distance $r_{ij}$ anymore, but actually on the relation between $\mathbf{r}_{ij}$ and $\mathbf{u}_i$. Intuitively, one can appreciate this fact as a simple consequence of the different $ij$ pairwise distance changes depending on the direction of $\mathbf{u}_{i}$. For example, when $\mathbf{r}_{ij}\cdot\mathbf{u}_i=0$, $\left|\Delta\mathbf{r}_{ij}\right|=\left|\mathbf{r}^{\prime}_{ij}-\mathbf{r}_{ij}\right|\sim0$, and thus the diagonal SLC parameter is very small in the presence of weak spin-orbit coupling. However, this can also be formally analyzed in terms of the perturbative (LKAG-like) expression for the calculation of the general term $\frac{\partial\mathcal{J}_{ij}^{\alpha\beta}}{\partial u_k^{\mu}}$ \cite{Mankovsky2022}: the analog of $\boldsymbol{\delta}_i$ (Eq. \ref{eq:lkag}) is a displacement matrix (linearized in $u_k$) whose elements are proportional to the spherical harmonics $Y_{lm}(\hat{\mathbf{u}}_k)$ in LMTO and the so-called Gaunt coefficients \cite{Gaunt1929}. Thus, while isotropic character of the Heisenberg exchange parameters (Eq. \ref{eq:lkag}) is essentially linked to the fact that $\boldsymbol{\delta}_i$ is a diagonal operator, the spin-lattice couplings will naturally depend on the $\mathbf{u}_k$ direction.

\begin{figure*}[!htb]
\includegraphics[width=\textwidth]{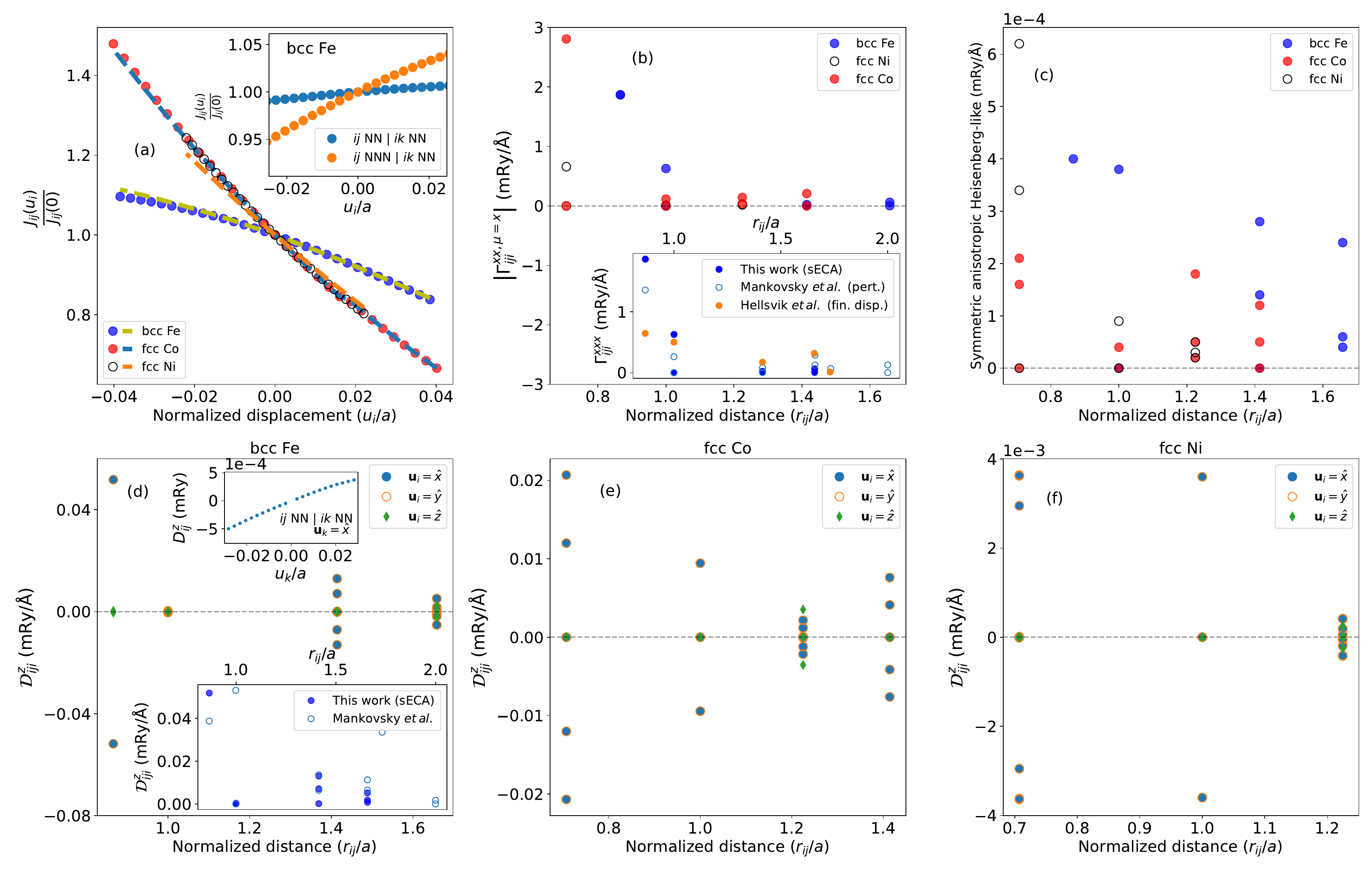}
\caption{(a) Normalized change of nearest-neighbor $ij$ pairwise interaction for Fe, Co and Ni as a function of the displacement $\left|\mathbf{u}_{i}\right|$ (in units of the lattice parameter). Points represent the values calculated with the simplified ECA scheme (sECA), while the dashed lines denote the calculated values using a fully self-consistent ECA scheme. \textit{Inset:} Variation of the Heisenberg exchange $J_{ij}$ in bcc Fe for $k\neq \{i,j\}$, considering $ij$ a pair of nearest-neighbors or next nearest-neighbors sites, while the displaced atom $k$ is positioned as a nearest-neighbor of site $i$. The displacement direction ${\mathbf{u}}_k$ is taken as ${\mathbf{u}}_k\parallel\mathbf{r}_{ik}$. (b) Comparison between the calculated diagonal SLC parameters of bcc Fe using sECA (see Section \ref{sec:eca}), as a function of $r_{ij}$, and the ones reported in Refs. \cite{Mankovsky2022,Hellsvik2019}, obtained using a perturbative approach. In this case, the displacement is taken in the $\hat{x}$ direction ($[100]$), while the spin quantization axis is in $\hat{z}$ ($[001]$). (c) Symmetric anisotropic Heisenberg-like interactions induced by the displacement of the $i$-th atom. (d-f) Same as in (c), but for the DM-like couplings for each $3d$ ferromagnet when spin-orbit is enabled. Negative displacements ($\left|\mathbf{u}_{i}\right|<0$) represent a reduction of the $ij$ pairwise distance, in real space. The calculated equilibrium nearest-neighbor exchange interactions are $J_{ij}(0)=1.31$ mRy for bcc Fe, $J_{ij}(0)=0.206$ mRy for fcc Ni, and $J_{ij}(0)=0.985$ mRy for fcc Co.}
\label{fig:mml-ferromagnets}
\end{figure*}

In turn, in Fig. \ref{fig:mml-ferromagnets}(a) an excellent agreement between the ECA (fully self-consistent) and the sECA is seen, evidencing the perturbative nature of the change in the electronic density for realistic displacements. Explicity, for $\Gamma_{iji}^{\alpha\alpha\mu}$ with $ij$ representing nearest-neighbors sites, the variation between the fully self-consistent and simplified schemes is $\sim3\%$ (bcc Fe), $\sim5\%$ (fcc Ni), and $\sim1.5\%$ (fcc Co). In turn, in Fig. \ref{fig:mml-ferromagnets}(a), \textit{Inset}, we demonstrate the effect provoked by a displacement of the atom $k\neq i(j)$ in the diagonal SLC parameters, for first and second nearest-neighbor $ij$ interactions, where a somewhat smaller variation from the equilibrium values can be observed. Due to their large multiplicity, however, this does not discard their role on the exchange-mediated magnetoelastic parameters, which will be further discussed in Section \ref{sec:connection-measurements}.

From the results displayed in Fig. \ref{fig:mml-ferromagnets}(b), among the three ferromagnets, fcc Co has the highest SLC parameters, and the fundamental reason for that (related to the Bethe-Slater behavior) will be explored futher on. The normalized variations shown in Fig. \ref{fig:mml-ferromagnets}(a) suggests that there is a strong connection between $\Gamma_{iji}^{\alpha\alpha\mu}$ with the bulk structure in real-space -- immediately noticed from the very close normalized variations on $J_{ij}$ for fcc Co and fcc Ni. Figure \ref{fig:mml-ferromagnets}(b) also shows the moderate decay rate of the diagonal part of $\Gamma_{ijk}^{\alpha\beta\mu}$ tensor for Fe, Co and Ni, as a consequence of the well-known itinerant character of magnetism in these materials. Formally, we can interpret this decay rate in a similar fashion as analyzed in Ref. \cite{Pajda2001}: for large separation distances $r_{ij}$, the Green's functions $\mathbf{G}_{ij}$ behaves proportional to $\frac{1}{r_{ij}}$, which leads to a general profile $\propto\frac{1}{r_{ij}^4}$ for the three-body interaction.

\begin{figure}[!htb]
\includegraphics[width=\columnwidth]{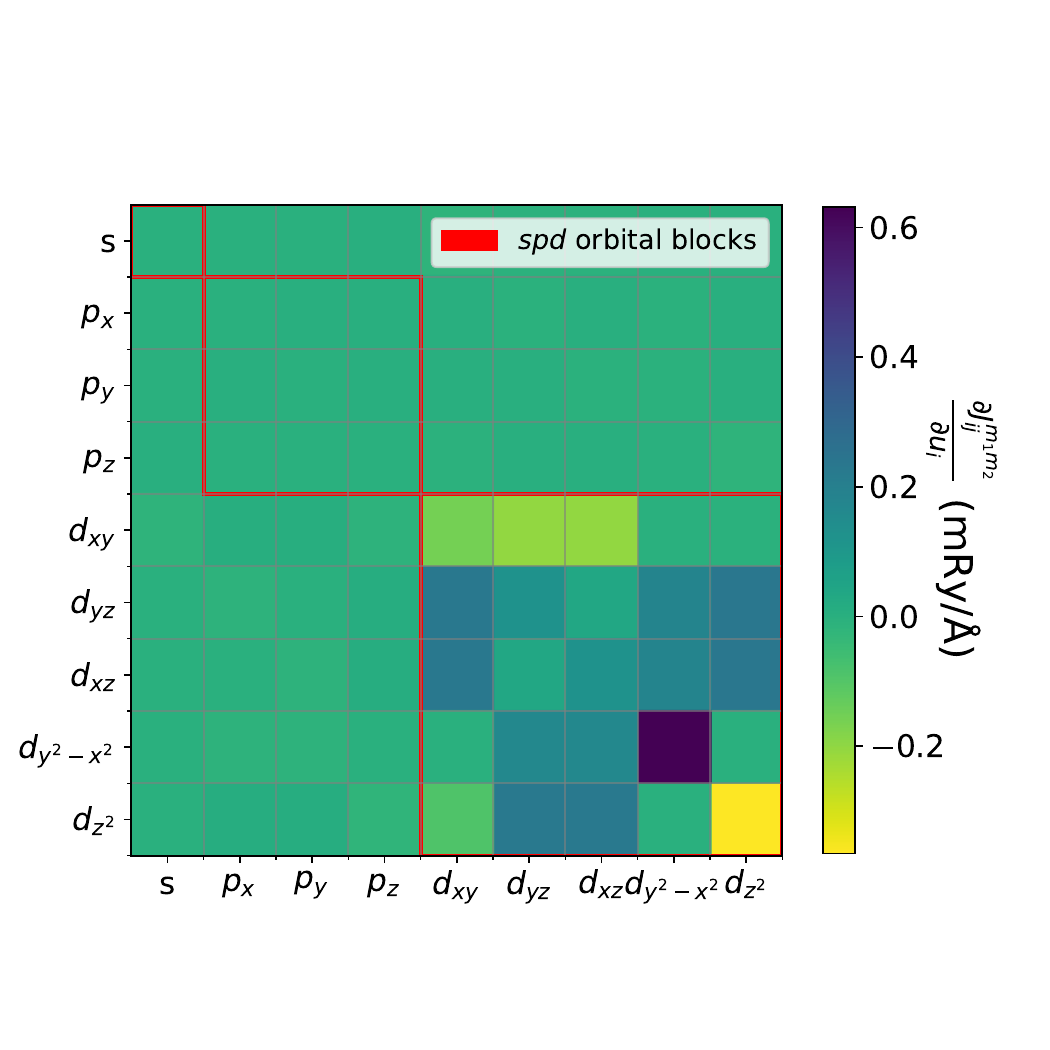}
\caption{Orbitally-decomposed nearest-neighbor diagonal SLC (Heisenberg-like) parameter of bcc Fe. The spin quantization axis is in $\hat{z}$ ($[001]$).}
\label{fig:orb-1st-fe}
\end{figure}

Concerning the mechanisms which give rise to the exchange interactions, an interesting observation was reported in Ref. \cite{Kvashnin2016}. The authors noticed that, among the $3d$ series, bcc Fe presents a quite unexpected and large antiferromagnetic contribution coming from the $t_{2g}$-$t_{2g}$ channel, which is also the main responsible for the RKKY-type oscillations. In that sense, as we are also dealing here with the same material, it is valuable to analyze the orbital decomposition of the spin-lattice couplings. Fortunately, in a finite-displacement approach (ECA or sECA), where in the limit of $u_k^\mu\rightarrow0$, $\Gamma_{ijk}^{\alpha\beta\mu}=\frac{\partial\mathcal{J}_{ij}^{\alpha\beta}}{\partial u_k^{\mu}}\approx\frac{\mathcal{J}_{ij}^{\alpha\beta}(u_k^\mu)-\mathcal{J}_{ij}^{\alpha\beta}(0)}{u_k^\mu}$, the Heisenberg-like spin-lattice coupling between the $m_1$-th orbital at site $i$ and the $m_2$-th orbital at site $j$ becomes simply

\begin{equation}
\label{eq:spin-lattice-orbital-decomposition}
\left(\Gamma_{ijk}^{\alpha\alpha\mu}\right)_{m_{1}m_{2}}\approx \frac{J_{ij}^{m_{1}m_{2}}(u_k^\mu)-J_{ij}^{m_{1}m_{2}}(0)}{u_k^\mu},
\end{equation}

\noindent in which $J_{ij}\equiv$ (Eq. \ref{eq:lkag}) is defined as $J_{ij}=\sum_{m_{1}m_{2}}J_{ij}^{m_{1}m_{2}}$ (with $J_{ij}^{m_{1}m_{2}}=\frac{1}{4\pi}\textnormal{Im}\int_{-\infty}^{E_F}\delta_{i}^{m_{1}}G_{ij}^{m_{1}m_{2},\uparrow}\delta_{j}^{m_{2}}G_{ji}^{m_{2}m_{1},\downarrow}d\epsilon$) \cite{Kvashnin2016,Korotin2015,Cardias2017}), because this decomposition is suitable for the non-relativistic level of theory. Thus, from Eq. \ref{eq:spin-lattice-orbital-decomposition} it is trivial that $\Gamma_{ijk}^{\alpha\alpha\mu}=\sum_{m_{1}m_{2}}\left(\Gamma_{ijk}^{\alpha\alpha\mu}\right)_{m_{1}m_{2}}$. As an example, the results for Fe are shown in Figure \ref{fig:orb-1st-fe}, where we consider only the nearest-neighbors interactions in the $k=i$ case.

First, as expected, one notices the (almost isolated) relevance of the $3d$-block to $\Gamma_{iji}^{\alpha\alpha\mu}$. However, differently from the situation of the lattice in its equilibrium structure ("\textit{static}" case) \cite{Kvashnin2016}, where all three orbital contributions ($e_{g}$-$e_{g}$, $e_{g}$-$t_{2g}$, and $t_{2g}$-$t_{2g}$) actively participate in generating the final value of the nearest-neighbor exchange interactions, here the short-ranged mixed ($e_{g}$-$t_{2g}$) couplings are the leading responsible for the spin-lattice, accounting for $\sim75\%$ of the final value. Note that $e_{g}$-$e_{g}$ presents the largest contributions that, nevertheless, compensate each other. As in Fe the Fermi surface (FS) has majorly a $d$-$t_{2g}$ character \cite{Schafer2005}, we can thus conclude that the nearest-neighbor spin-lattice couplings in this material are predominantly influenced by contributions from the so-called Fermi sea, with interactions mediated by the double-exchange mechanism \cite{Kvashnin2016}. On the other hand, further away from the first neighborhood, the $t_{2g}$-$t_{2g}$ coupling becomes more relevant (data not shown), highlighting the greater weight of contributions arising from RKKY-type interactions, directly linked to modifications in the topology of the FS induced by atomic displacements. Of course, as discussed before and differently from the first neighbors case, such contributions on longer-range interactions vary depending on the relation between $\mathbf{r}_{ij}$ and $\mathbf{u}_k$.

As in the static case, Fe appears to be a singular example. When the same analysis is applied to nearest neighbors couplings in fcc Co, for example, the mixed interactions show the least contributions to the final spin-lattice values. In the whole interaction matrix (analogous to Fig. \ref{fig:orb-1st-fe}, not shown), the existent negative contributions to the spin-lattice are much smaller ($>50\%$) than the calculated for Fe, in line with the orbitally-decomposed Bethe-Slater behavior shown by Cardias \textit{et al.} \cite{Cardias2017}.

\subsubsection{Parameters with spin-orbit coupling origin}

When spin-orbit is taken into account, Dzyalonshinskii-Moriya-like (DM-like, $\boldsymbol{\mathcal{D}}_{ijk}^{\mu}$) couplings can emerge with the displacements of the atoms from their equilibrium positions. In all $3d$ ferromagnets analyzed here, the DM is trivially zero due to the crystal symmetries in equilibrium, following Moriya's rules \cite{Moriya1960}. However, when an atom is displaced, the inversion symmetry is locally broken, allowing for the emergence of the anisotropic DM interaction; interestingly, those DM couplings have a \textit{dynamical} nature. The calculated $\mathcal{D}_{iji}^{z,\mu}$ component values, defined as $\mathcal{D}_{iji}^{z,\mu}=\frac{1}{2}(\Gamma_{iji}^{xy\mu}-\Gamma_{iji}^{yx\mu})$ \cite{Udvardi2003,Borisov2023}\footnote{Generally, $\mathcal{D}_{ijk}^{\gamma,\mu}=\frac{1}{2}\varepsilon_{\alpha\beta\gamma}(\Gamma_{ijk}^{\alpha\beta\mu}-\Gamma_{ijk}^{\beta\alpha\mu})$, where $\varepsilon_{\alpha\beta\gamma}$ is the Levi-Civita symbol.}, are shown in Figs. \ref{fig:mml-ferromagnets}(d-f) for $\mu\in\{x,y,z\}$ as a function of the $ij$ pair distance. Due to the relatively low spin-orbit strength of Fe, Ni and Co (with self-consistent values $\xi_{\textnormal{Fe}}^{\uparrow}\sim60$ meV, $\xi_{\textnormal{Fe}}^{\downarrow}\sim47$ meV; $\xi_{\textnormal{Ni}}^{\uparrow}\sim93$ meV, $\xi_{\textnormal{Ni}}^{\downarrow}\sim89$ meV; $\xi_{\textnormal{Co}}^{\uparrow}\sim76$ meV, $\xi_{\textnormal{Co}}^{\downarrow}\sim66$ meV), the $\hat{z}$-component of the emergent DM-like spin-lattice couplings is about two orders of magnitude (or more) lower than the Heisenberg-like elements, in meV/\AA. As argued in Refs. \cite{Mankovsky2022,Lange2023,Mankovsky2023}, these DM-like terms act as a channel for angular momentum transfer between the spins and the lattice subsystems, and may be of particular relevance in the understanding of ultrafast demagnetization experiments \cite{Tauchert2022}. 

Differently from Heisenberg-like interactions (diagonal parts), an important property of the DM-like vectors is the intrinsic relation of not only their magnitude, but also their direction with the ionic displacement $\mathbf{u}_{i}$. For example, when calculating the $\hat{z}$-component of $\boldsymbol{\mathcal{D}}_{ijk}^{\mu}$, a displacement perpendicular to that axis (say $\mathbf{u}_{i}=\hat{x}$ or $\mathbf{u}_{i}=\hat{y}$) produces the values shown in Fig. \ref{fig:mml-ferromagnets}(d), which are in generally good agreement with those reported in Ref. \cite{Mankovsky2022}\footnote{Small differences may arise due to the use of distinct methods to calculate the spin-lattice interactions (namely, the \textit{perturbative} approach in Ref. \cite{Mankovsky2022}, and the sECA in the present article).}. Of course, in this case the positive or negative signs are associated with the relative orientation of that displacement. However, when the displacement is also in $\hat{z}$, $\mathcal{D}_{iji}^{z,\mu=z}$ becomes virtually zero for nearest to third neighbors, acquiring back residual values for the fourth neighboring shell (comparable to the ones obtained for $\mathbf{u}_{i}=\hat{x}$ or $\mathbf{u}_{i}=\hat{y}$). Analogous (but not exactly) is also the case for fcc Ni and fcc Co. Note that, even though we show results for the $\hat{z}$-component of $\boldsymbol{\mathcal{D}}_{ijk}^{\mu}$, the two remaining components can be fully determined by cubic symmetry transformations \cite{Lange2023}.

Physically, this again can be understood in terms of Moriya's rules. On the one hand, taking $ij$ as a nearest-neighbor pair in bcc Fe, any $\mathbf{u}_{i}=\hat{z}$ do not violate the mirror plane encompassing the line segment between $i$ and $j$. Consequently, this configuration is expected to give rise to DM-like interactions in a direction perpendicular to this plane, or parallel to the $xy$-plane ($\mathcal{D}_{iji}^{z,\mu=z}\rightarrow0$). On the other hand, if $ij$ is now a fourth-neighbor pair, some $\mathbf{r}_{ij}$ vectors are contained in planes that do not constitute an inversion mirror (e.g., the $(\bar{1}\bar{3}0)$ plane), so in principle non-zero $\mathcal{D}_{iji}^{z,\mu=z}$ are allowed; in this particular case, two different non-zero values are predicted, because the ions that live in such planes are positioned in two symmetric heights w.r.t. the $xy$-plane (e.g., $(-\frac{3}{2}a,\frac{1}{2}a,\pm\frac{1}{2}a)$). Comparable reasoning can be drawn for the other $3d$ ferromagnets. Thus, the interconnection between the system's symmetry, distance of the $ij$ pair, and chemical classification of the involved atoms can generate very complex and anisotropic (within each neighboring shell) patterns of the spin-lattice DM-like couplings, resembling another spin-orbit-related quantity: the Gilbert damping \cite{Lu2023}. Formally, the anisotropic behaviors of both quantities are explained by the requirement of non-diagonal operators, although with different origins: while the former is related to a non-local displacement matrix $\left(\propto\frac{\partial\mathcal{H}}{\partial\mathbf{u}}\right)$, the latter comes, in the torque-correlation model, from a local spin-orbit torque operator \cite{Lu2023,Thonig2018}.

For the DM-like terms when $k\neq i(j)$, analogous reasoning can be drawn as for the isotropic contributions, respecting the general symmetry properties of the $\boldsymbol{\mathcal{D}}_{ijk}^{\mu}$ parameters discussed above. The norm becomes at least one order of magnitude lower when compared to the $k = i$ counterpart situation, with its $\mu$-component approaching zero whenever the $ij$ pair belongs to a mirror plane in $\mu$ (which is not violated by the displacement $\mathbf{u}_{k\neq i(j)}$). An example of the calculated values of $D_{ij}^{z}$ for $k \neq i(j)$ (with $ij$ and $ik$ pairs of nearest-neighbor sites) in bcc Fe is shown in the inset of Fig. \ref{fig:mml-ferromagnets}(d), as a function of the displacement. For this case, we obtain a variation of $\mathcal{D}_{ijk}^{z,\mu=z}\sim 0.005$ mRy/\AA for the atoms $i,j,k$ at $(\frac{1}{2}a,\frac{1}{2}a,\frac{1}{2}a)$, $(a,0,0)$, and $(0,0,0)$, respectively -- however $\mathcal{D}_{ijk}^{z,\mu=z}\rightarrow0$ if the atom $j$ is positioned at $(0,0,a)$ for $\textbf{u}_{k}=\hat{z}$. Although generally the case for pure ferromagnets, the order-of-magnitude difference between $k\neq i(j)$ and $k = i$ couplings may be not true when the displaced atom $k$ is a high spin-orbit center following the Fert-Levy model for the DM interactions \cite{Fert1980}, as also reported by Ref. \cite{Lange2023}. Natural examples are L1$_0$ FePt, CoPt, P$m\bar{3}m$ FeRh, among others. In such materials, it is expected that the magnetoelastic Hamiltonian in momentum space (to describe the magnon-phonon interactions) \cite{Ruckriegel2020} will require accounting for several neighbor combinations in real space during the Fourier transformations due to the relevance of the long-range DMI-like couplings.

Together with the DMI-like terms, naturally also symmetric-anisotropic Heisenberg interactions will be induced by the local breaking of inversion symmetry when an atom is displaced; the $\hat{z}$-component, for instance, can be calculated as $\varGamma_{ijk}^{z,\mu}=\frac{1}{2}(\Gamma_{ijk}^{xy\mu}+\Gamma_{ijk}^{yx\mu})$. As can be seen in Fig. \ref{fig:mml-ferromagnets}(c), they are at least one order of magnitude lower than the correspondent antisymmetric (DMI-like) interactions -- a trend that is thus followed by its contribution to the magnetoelastic energy density in the continuum limit (see Appendix \ref{sec:continuum-limit}). Moreover, from the $3d$ ferromagnets analyzed, surprisingly Ni has the strongest induced $\varGamma_{ijk}^{z,\mu}$ interactions. When emergent in a static lattice primarily due to geometric reasons, as in the case of the kagome $A$V$_3$Sb$_5$ (where $A$ is either Cs, Rb, K) compounds \cite{Karmakar2023,Hasan2023}, the symmetric anisotropic Heisenberg exchange couplings have been argued to be relevant to stabilize complex magnetic textures \cite{Karmakar2023,Yambe2022} and nonreciprocal magnons \cite{Hayami2022}. Thus, when those terms arise from the coupling to the lattice degrees of freedom, they might be linked to the emergence of nonreciprocal magnetoacoustic waves.

\subsubsection{Comparison with the fitting procedure}

\begin{figure}[!htb]
\includegraphics[width=0.5\textwidth]{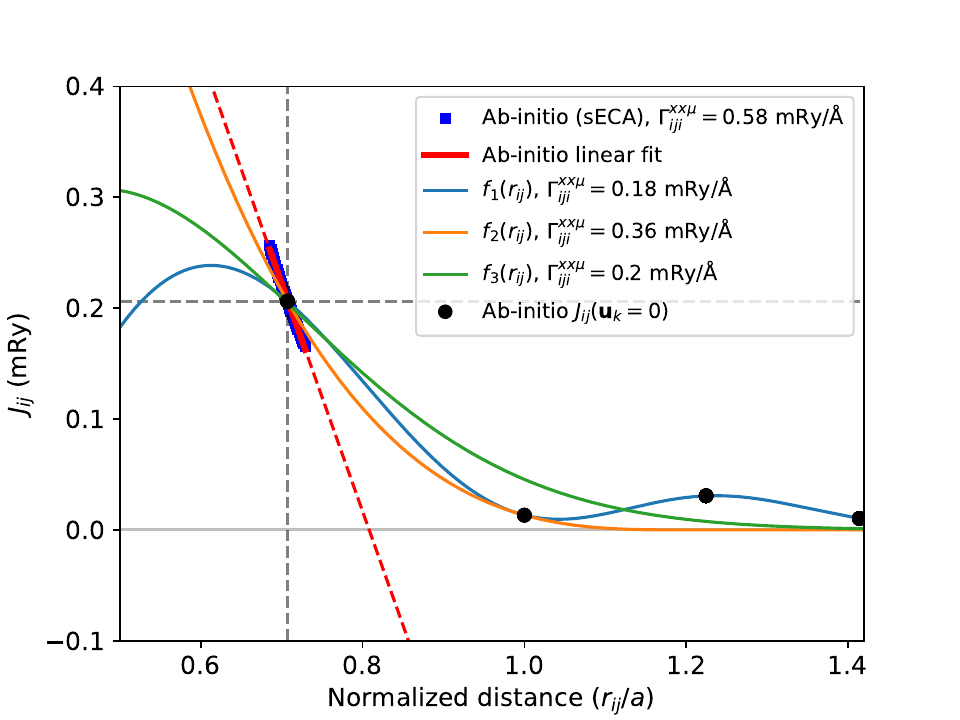}
\caption{First-principles calculated exchange interactions $J_{ij}$ (black circle dots) of fcc Ni and the correspondent change (blue square dots) when the atom $i$ in the nearest-neighbor $ij$ pair is displaced, as a function of the normalized interatomic distances ($r_{ij}/a$). The curves represent the best fit from the functions defined in Eq. \ref{eq:modelsjij}. The Heisenberg-like parameters (diagonal elements) $\Gamma_{iji}^{xx\mu}=\frac{\partial J_{ij}^{xx}}{\partial u_i^{\mu}}$ (for $\mu=\hat{x}$) obtained for the nearest-neighbors in each case are shown in the legend.}
\label{fig:comparison-fitting}
\end{figure}

Here we shall briefly investigate the fitting procedure described in Section \ref{sec:determination-by-fitting}. To better understand the difference with respect to directly calculated parameters by first-principles methods (such as, e.g., in Refs. \cite{Hellsvik2019,Mankovsky2022,Sadhukhan2022}), we can compare the \textit{ab-initio} $\Gamma_{ijk}^{\alpha\beta\mu}$ values with the curves predicted by both the BS and the damped-oscillatory models within a realistic range of atomic displacements (see Appendix \ref{sec:realistic-atomic-displacements}).  An example for fcc Ni is shown in Fig. \ref{fig:comparison-fitting}, comparing only the diagonal (Heisenberg-like) spin-lattice couplings, namely $\Gamma_{iji}^{xx\mu}$ (i.e., with $i=k$). The obtained fitting parameters are in agreement with those available in literature \footnote{The fitting parameters used are (with [$\ldots$] denoting their units): for function $f_1$, $(J_0, C_1, C_2, d_1, d_2, \omega, \xi, \gamma)=$ (0.28 [mRy], 1.61, 0.27, 0.67 [$a$], 1.20 [$a$], 5.29 [$1/a$], 3.86 [$1/a$], 23.33 [$1/a^2$]), see Ref. \cite{Amann2018}; for function $f_2$, $(J_0, r_c)=$ (3.02 [mRy], 1.2 [$a$]); and, finally, for function $f_3$, $(\alpha_J, \delta_J, \gamma_n, r_c)=$ (0.21 [mRy], 0.48 [$a$], $4.1\times10^{-5}$, 1.42 [$a$])}, see Ref. \cite{Tranchida2018}.

As can be seen from Fig. \ref{fig:comparison-fitting}, the $\Gamma_{iji}^{xx\mu}$ values can vary significantly with respect to those calculated via the sECA method (of the order of $\sim70\%$ in the worst case -- function $f_1$), with a tendency to underestimation. A possible explanation relies on the inadequacy of the adiabatic approximation for Ni \cite{Kudrnovsky2008,Bruno2003} and, likely, the sensitivity of the fitting procedure with underestimations in the exchange parameters already on the next-nearest-neighborhood (i.e., even with small chosen $r_c$); therefore, in this case the fixed inter-atomic dependence assumed to $J_{ij}({r}_{ij})$ for small local displacements becomes less justified. 

In particular, for a comprehensive description of the interaction between spin waves and phonons, including spin-gradient terms and beyond, it is relevant to recall that the simplified treatment given by all fitting approaches via the functions defined in Section \ref{sec:determination-by-fitting} lacks at least two important aspects, as discussed previously: (\textit{i}) the spatial dependence of the $\mathbf{u}_k$ displacements, which encompasses information about the lattice structure; and (\textit{ii}) the absence of DMI-like and anisotropy-like terms, which would enable magnon-number nonconserving processes.

\subsection{Connection to macroscopic observables}
\label{sec:connection-measurements}

Even though the listed properties for $\Gamma_{ijk}^{\alpha\beta\mu}$ so far seem reasonable, one may wonder about the validity of such atomistic parameters in a macroscopic frame, and if they can be directly linked to experimental observations somehow. As discussed in Ref. \cite{Streib2019}, a simple quantity to which $\frac{\partial J_{ij}}{\partial u_k}$ can be related is the so-called magnetic Grüneisen parameter $\gamma_m$ \cite{Bloch1966}, accessible by experiments \cite{Kamilov1998}. This parameter is defined  as $\gamma_m=-\frac{\partial\ln T_C}{\partial\ln V}$. Assuming that the Curie temperature ($T_C$)
depends linearly on the exchange constants, as suggested by the mean-field approximation (MFA), we find $\gamma_m\sim-\frac{\partial\ln J_0}{\partial\ln V}=-\frac{J_0^{\prime}a}{3J_0}$. Here, $a$ denotes the equilibrium distance, $J_0=\sum_{j\neq0}J_{0j}$, and $J_0^{\prime}=\sum_{j\neq0}\frac{\partial J_{0j}}{\partial a}$ for a reference site at $i=0$. Despite the well-known limitations of the MFA, this approximation is still suitable (and commonly applied) to simple $3d$ ferromagnets such as bcc Fe \footnote{Even though data is also available for fcc Ni (see, for instance, Refs. \cite{Leger1972, Kamilov1998}), it is known that the exchange constants derived from the magnetic force theorem are optimal for calculations of the spin-wave stiffness $D$, rather than thermodynamical properties such as the Curie temperature \cite{Katsnelson2004}. Thus, the Heisenberg model parameters result in a $T_C$ for fcc Ni that is usually a factor $\sim\frac{1}{2}$ too small, also in the mean-field approximation picture \cite{Solovyev2021,Katsnelson2004,Pajda2001,Halilov1998}. In this sense, bcc Fe is a more suitable material for the analysis of $\gamma_m$ as in the main text.}; we can recognize this fact by comparing the moderate difference between the MFA and the more sophisticated Monte Carlo $T_C$ estimates for bcc Fe, using the here calculated $J_{ij}$'s: $1273$ K and $975$ K, respectively, both in fair agreement with the experimental value ($1044$ K). 

By combining the data for variation of $T_C$ induced by pressure in bcc Fe \cite{Leger1972} with the ordinary pressure $\leftrightarrow$ volume relation given by Vinet-type equation of state \cite{Dewaele2006}, we obtain the experimental value $\gamma_m^{\textnormal{Expt}}\sim-1.51$. From the theoretical side, adopting the reasonable $k=i$ premise for this case \footnote{We can be convinced by this claim as $\gamma_m$ is originally defined by the change in $T_C$ w.r.t. the volume variation. When calculating $\gamma_m$ directly from the pressure data shown in Appendix \ref{sec:appendix-pressure}, the mean-field approximation yields $\left|\gamma_m\right|\sim1.1$, or $J_0^{\prime}\sim12.9$ mRy/Å, reasonably close to our estimation from the $k=i$ assumption.} yields $J_0^{\prime}\sim17.5$ mRy/Å (up to the fourth neighboring shell) and $J_0\sim11.2$ mRy, leading to $\gamma_m^{\textnormal{Theo}}\sim-1.49$. This result is in good agreement with $\gamma_m^{\textnormal{Expt}}$, in spite of the approximations involved.

\begin{figure}[!htb]
\includegraphics[width=\columnwidth]{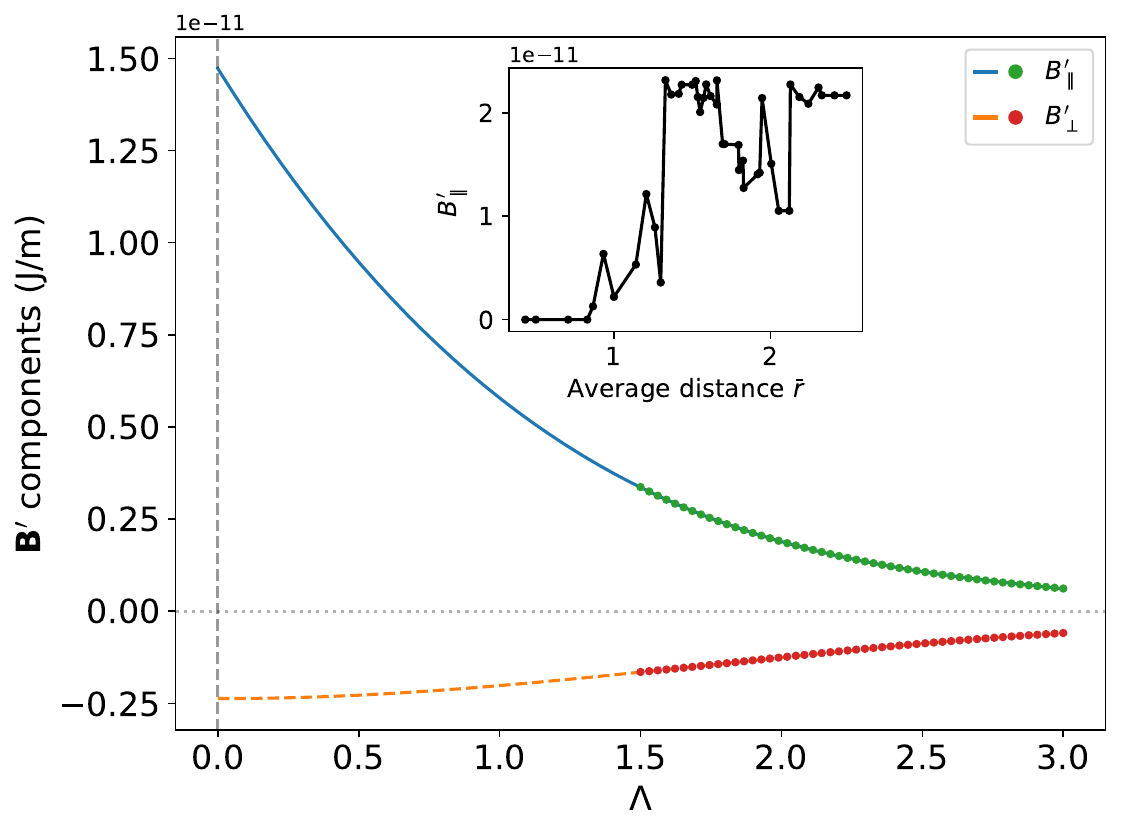}
\caption{$B^{\prime}_{\parallel}$ and $B^{\prime}_{\perp}$ for bcc Fe as a function of the attenuation constant $\Lambda$, extrapolated to $\Lambda\rightarrow0$. The lines indicate the polynomial function fit for each case. \textit{Inset:} the value of $B^{\prime}_{\parallel}$ as a function of $\bar{r}$ when $\Lambda=0$.}
\label{fig:b_prime_components}
\end{figure}

In addition to the magnetic Grüneisen parameter, the quantum-mechanically derived spin-lattice interactions are also strongly linked to classical magnetoelastic theory and the associated phenomena. A more detailed derivation into that direction is shown in Appendix \ref{sec:continuum-limit}, as well as the extended energy density expression, including terms that were not traditionally considered in early theoretical developments by Abrahams and Kittel (AK) \cite{Abrahams1952} and KT \cite{Kaganov1959}. If we momentously disregard the extra terms derived in Appendix \ref{sec:continuum-limit} and write $\varepsilon_{me}$ following the approach of KT, we have \cite{Streib2019,Kaganov1959,Abrahams1952,Kittel1949}

\begin{equation}
\begin{split}
\label{eq:magnetoelastic-energy2}
\varepsilon_{me}=\varepsilon_{ms}+\varepsilon_{ex}=\\
\int d\mathbf{r}\sum_{\mu\kappa}B_{\mu\kappa}m^{\mu}m^{\kappa}\epsilon_{\mu\kappa}+\int d\mathbf{r}\sum_{\mu\kappa}B^{\prime}_{\mu\kappa}\frac{\partial\mathbf{m}}{\partial x_{\mu}}\cdot\frac{\partial\mathbf{m}}{\partial x_{\kappa}}\epsilon_{\mu\kappa},
\end{split}
\end{equation}

\noindent where $\varepsilon_{ms}$ is the magnetostriction term (of relativistic origin), and $\varepsilon_{ex}$ is the term that describes the change in exchange energy due to strain. While the former has been connected with effects such as the \textit{magnetoelastic-gap} \cite{Turov1983} and damping of magnetization \cite{Streib2018} (which is not of the Gilbert type), the latter has been associated with phenomena such as the attenuation of sound waves \cite{Lord1968} (i.e., acting effectively as a source of damping to phonons, together with other sources, such as phonon-phonon scattering \cite{Maldonado2017}), and, conversely, also with damping of magnons \cite{Streib2019}. The link to the microscopic $\Gamma_{ijk}^{\alpha\beta\mu}$ parameters is thus given as (derived in Appendix \ref{sec:continuum-limit})

\begin{equation}
\begin{split}
\label{eq:tensor-b-elements}
B_{\mu\kappa}=\lim_{\Lambda\rightarrow0}\frac{1}{4V}\sum_{k}\left(\Gamma_{iik}^{\mu\kappa\mu}r_{ik}^{\kappa}+\Gamma_{iik}^{\mu\kappa\kappa}r_{ik}^{\mu}\right)e^{-\Lambda r_{ik}}\\
B^{\prime}_{\mu\kappa}=\lim_{\Lambda\rightarrow0}\frac{1}{8V}\sum_{\alpha jk}\left(\Gamma_{ijk}^{\alpha\alpha\mu}r_{ij}^{\mu}r_{ij}^{\kappa}r_{ik}^{\kappa}+\Gamma_{ijk}^{\alpha\alpha\kappa}r_{ij}^{\kappa}r_{ij}^{\mu}r_{ik}^{\mu}\right)e^{-\Lambda\bar{r}},
\end{split}
\end{equation}

\noindent in which $V$ is the volume of the unit cell, $\Lambda$ is a non-negative attenuation constant, and $\bar{r}=\frac{1}{2}\left(\left|\mathbf{r}_{ij}\right|+\left|\mathbf{r}_{ik}\right|\right)$ is an average distance. Particularly for the cubic symmetry the tensors elements $B_{\mu\kappa}$ and $B^{\prime}_{\mu\kappa}$ reduce to $B_{\mu\kappa}^{(\prime)}=B_{\parallel}^{(\prime)}\delta_{\mu\kappa}+B_{\perp}^{(\prime)}(1-\delta_{\mu\kappa})$, which is also obtained numerically here. While the first term of Eq. \ref{eq:tensor-b-elements} just partially new in literature (as it has been evaluated via \textit{ab-initio} before using distinct approaches, as the total energy or torque method calculations \cite{Burkert2004,Fanhle2002}), the second term is still less explored. With this motivation in hand, Figure \ref{fig:b_prime_components} shows the convergence of both tensor elements from Eq. \ref{eq:tensor-b-elements} as a function of $\Lambda$ for bcc Fe, considering all 3-body interactions up to the fourth neighboring shell from the reference $i$-th atom. The limit $\Lambda\rightarrow0$ was obtained by an extrapolation method via a third order polynomial fit. From the \textit{Inset}, we first notice the oscillating character of $B^{\prime}_{\parallel}$ as a function of distance, a direct consequence of the long-range RKKY behavior. As for the contributions, $k=i$ is found to be always negative, while the $k\neq i(j)$ terms, due to their multiplicity, were found to be relevant for obtaining specially the diagonal elements of the $\mathbf{B}^{\prime}$ tensor: they lead, for instance, to the final positive value of $B_{\parallel}^{\prime}$.

\subsection{Noncollinearity: dependence on the average spin polar angle $\bar{\theta}$ and connection to spin temperature}
\label{sec:temperature}

In the context of magnetization switching, ultrafast demagnetization \cite{Windsor2021}, magnetocalorics, skyrmion transport \cite{Wu2021}, and several other phenomena in which the spin-lattice coupling plays a role, thermal effects are commonly present. Associated with them, the spin structure experiences a noncollinear ordering. Thus, still in the same context to evaluate the $ij$-mediated interactions, an analysis on how the SLC parameters behave when the atomistic spins are driven to a noncollinear configuration is desirable from the theory point of view.

The direct connection with the temperature of the spin subsystem ($T_s$) is complex. As noted by several authors (see the reviews  \cite{Borisov2023,Szilva2023} and references therein), the scheme designed by Liechtenstein \textit{et al.} \cite{Liechtenstein1987} to calculate exchange interaction parameters depends on a \textit{reference state}, which is usually taken to be the spin texture that represents the lowest energy and was originally defined to be ferromagnetic (the case, e.g., for Fe, Co, Ni, and Gd). However, for \textit{each} excited noncollinear spin configuration dynamically driven by thermal fluctuations there exists a set of interaction parameters that defines a local (and temporary) mapping of the spin Hamiltonian (Eq.  \ref{eq:spin-lattice-hamiltonian}). In this sense, a route towards a general solution would be to consider an extension of the approach reported in Ref. \cite{Stockem2018}: a coupled constrained noncollinear DFT-atomistic spin dynamics (ASD) method, which still relies on an adiabatic approximation of the spin degrees of freedom with respect to the electronic structure. In addition of being computationally expensive, this approach would involve great complications in the interpretation of numerical results, given the simultaneous dependence of the spin-lattice parameters on $\textbf{r}_{ij}$, $\textbf{u}_k$, and the spin state itself. 

To facilitate our analysis, we can largely simplify the problem and fit it into the framework of Ref. \cite{Rodrigues2016}. In that study, the authors considered that there exists a correspondence between $T_s$ and the average spin moment $\bar{\theta}$ coordinate, for fixed or randomly distributed $\phi$ values. This assumption produced a good comparison between experimental and theoretical magnon spectra when the correspondent non-collinear Heisenberg parameters $\left\{J_{ij}^{\textnormal{NC}}(\bar{\theta})\right\}$ were used, instead of the usual dataset (i.e., that considers the ferromagnetic configuration as the reference state). Given the success of this simplified approach, we also consider the direct $T_s\leftrightarrow\bar{\theta}$ correspondence here.

In steps of $\Delta T_s=3$ K, we calculated the average $\bar{\theta}$ for each $T_s$, also exceptionally taking into account the cubic MAE with experimental constants (see Appendix \ref{sec:theta-bar-appendix}). This condition ensures the existence of a preferential axis and, consequently, removes the ambiguity in defining the same reference Cartesian $\{\hat{x},\hat{y},\hat{z}\}$ axes for all independent calculations (as, in principle, without MAE and for $\left|\mathbf{D}_{ij}\right|=0$, Eq.  \ref{eq:generalized-bilinar-hamiltonian} is rotationally invariant). The calculated $\bar{\theta}$ from the smoothed functions for given low spin temperatures are presented in Table \ref{tab:theta-averages}. Interestingly, the angles $\bar{\theta}$ for bcc Fe are close to those found in Ref. \cite{Rodrigues2016} for six layers of Fe(001) on top of Ir(001), although slightly smaller for the same spin temperature (correctly indicating here the tendency towards a higher transition temperature due to the enhanced dimensionality), and reflecting distinct atomic structures (bcc \textit{versus} fcc in Ref. \cite{Rodrigues2016}).

\begin{table}[ht!]
\caption{Average spin moment polar angle $\bar{\theta}$, obtained from the smoothed functions shown in Fig. \ref{fig:theta-temperature} for selected low spin temperatures $T_s$.}
\begin{center}
\label{tab:theta-averages}
\begin{tabular}{c c c c c c}
\hline 
\hline
&&& \\[-2.5mm] 
& 20 K  & 40 K  & 60 K & 80 K & 100 K
\\ 
&&& \\[-3mm] \hline
&& &\\[-3mm] 	    
bcc Fe   & $5^{\circ}$	& $7^{\circ}$ & $9^{\circ}$ & $11^{\circ}$ &  $14^{\circ}$ \\
&&& \\[-3mm] 
fcc Co    & $5^{\circ}$	& $7^{\circ}$  & $10^{\circ}$ & $11^{\circ}$ & $13^{\circ}$ \\ 
&&& \\[-3mm] 
fcc Ni   & $6^{\circ}$ 	& $10^{\circ}$ &  $13^{\circ}$  & $15^{\circ}$ &  $17^{\circ}$
\\\hline\hline
    \end{tabular}
 \end{center}	
\end{table}

\begin{figure}[!htb]
\begin{center}
\includegraphics[width=0.7\columnwidth]{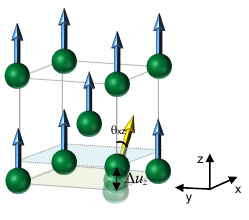}
\caption{Schematic representation of coupled degrees of freedom in an ideal bcc lattice: atomic displacement in the $\mu=\hat{z}$ direction and single-spin rotation by a polar angle of $\bar{\theta}$ in the $\hat{x}$-$\hat{z}$ plane, mimicking the noncollinearity induced by a finite spin temperature. Analogous calculation procedure is adopted for displacements in $\mu=\hat{x},\hat{y}$ directions.}
\label{fig:schematic-noncol-disp}
\end{center}
\end{figure}

\begin{figure*}[ht!]
    \centering
    \includegraphics[width=2.5\columnwidth]{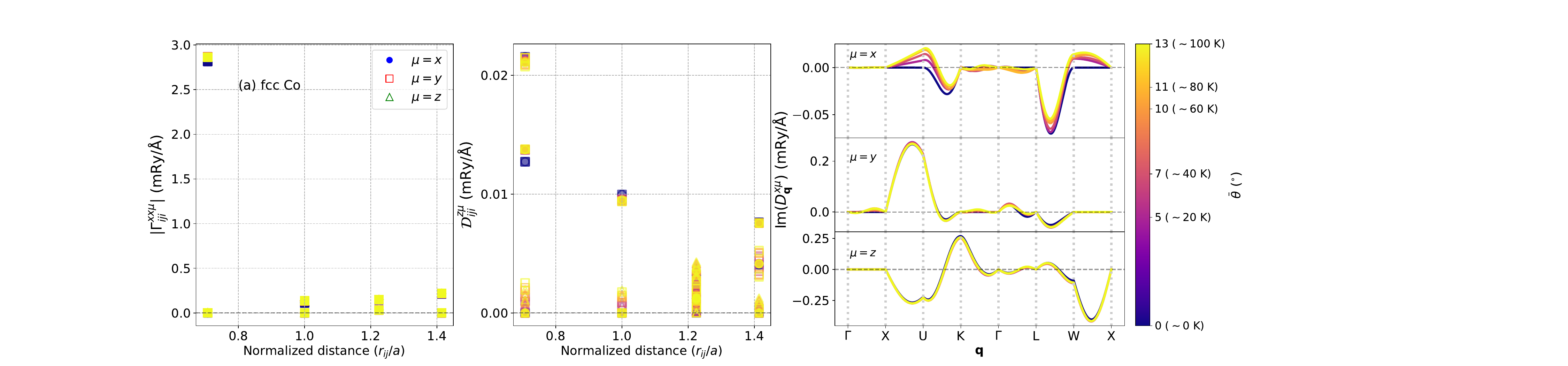} \\
    \vspace{-0.4em} 
    \includegraphics[width=2.5\columnwidth]{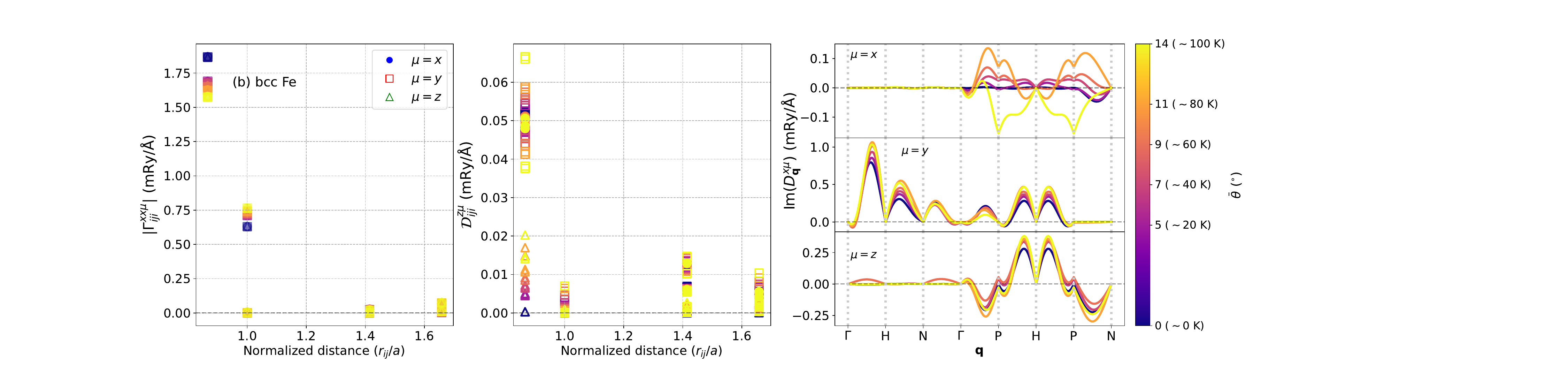} \\
    \vspace{-0.4em} 
    \includegraphics[width=2.5\columnwidth]{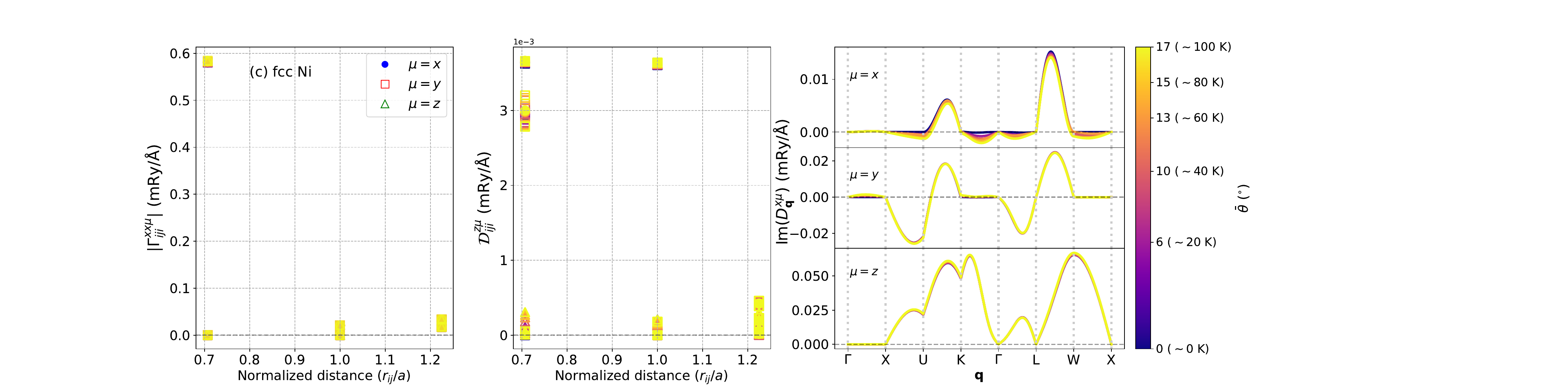}
    \caption{Heisenberg-like (left columns) and DMI-like (central  columns) spin-lattice parameters and the imaginary part of $\mathcal{D}_{\mathbf{q}}^{x\mu}$ (right columns) as a function of the polar angle $\bar{\theta}$ of the reference spin, which can be related to the spin temperature under a simplified picture: (a) fcc Co (topmost row); (b) bcc Fe (central row); (c) fcc Ni (bottom row).}
    \label{fig:spin-lattice-temperature}
\end{figure*}

Now that we established the rough $T_s\leftrightarrow\bar{\theta}$ connection, in order to evaluate the SLC parameters in that noncollinear environment, we use the general formalism by Fransson \textit{et al.} \cite{Fransson2017} (see Methods) together with the sECA scheme; the agreement between the sECA and the fully self-consistent ECA method is satisfactory in all cases. Here, the noncollinear calculations rely on an approximate constraining approach \footnote{We perform noncollinear calculations as implemented in RS-LMTO-ASA (see Ref. \cite{Bergman2007} for details), where the spin quantization axis (SQA) of the central atom is rotated by the polar angle $\bar{\theta}$, while the SQAs of the other atoms are kept fixed in the $\hat{z}$ direction.}, which is expected to produce good electronic structure results for small angles in elemental ferromagnets, such as those in Table \ref{tab:theta-averages}, in comparison to the more robust and theoretically well grounded constraining field DFT method \cite{Singer2005,Dederichs1984}. Figures \ref{fig:schematic-noncol-disp} and \ref{fig:spin-lattice-temperature} illustrate, respectively, the schematic of coupled degrees of freedom (atomic displacements and spin rotation), and the results for both Heisenberg-like and DM-like contributions for our $3d$ ferromagnets reference systems.    

Even though the limitations of our current procedure prevent achieving higher spin temperatures as we would have liked \footnote{Besides the validity of the approximate constraining approach, the low temperatures also justify the linear approximation for the Holstein-Primakoff transformations, in which we assume that the number of perturbations about the classical ground-state is very small.} (and as are routinely considered in experiments, usually at room-temperature \cite{Tauchert2022,Windsor2021}), we can already identify some relevant trends. From Figure \ref{fig:spin-lattice-temperature}, we see that fcc Co Heisenberg-like interactions, in contrast of being characterized by the largest spin-lattice parameters, present an almost negligible change with respect to the polar angle $\bar{\theta}$, in compliance with earlier results, that classified this system as a quasi-ideal Heisenberg magnet \cite{Chimata2017} -- i.e., with exchange couplings that are poorly configurational dependent. This is also followed by the DMI-like interactions. Maintaining the same pattern as Co, fcc Ni is even less affected by the single-spin rotation. The constancy of Co and Ni $J_{ij}$ under noncollinearity can be understood by a simultaneous decrease of the magnetic moments and increase of the exchange integrals; this stability of $J_{ij}(\bar{\theta})$ leads to the coupling with the lattice degrees of freedom being closely analogous to that of the reference ferromagnetic state ($\bar{\theta}=0$). In turn for bcc Fe, the single-spin rotation by $\bar{\theta}$ provokes a large change of the spin-lattice parameters, mainly influenced by the non-Heisenberg character of the $e_g$-$t_{2g}$ and $e_g$-$e_g$ channels \cite{Szilva2017}, whose short-ranged interactions are strongly driven by the double exchange mechanism \cite{Kvashnin2016}. As in the case of fcc Co, this higher configurational dependence is also reflected on the DMI-like couplings. In light of connections to damping \cite{Hong2024} and energy/angular momentum transfer to the lattice subsystem \cite{Mankovsky2022}, those results initially suggest that the effects/contributions associated with spin-lattice coupling in bcc Fe become more prominent with an increase in temperature, while on fcc Co they tend to be less sensitive -- something that will be investigated further. This observation builds upon the recent findings of Mankovsky \textit{et al.} \cite{Mankovsky2024}, who examined the influence of the \textit{lattice} temperature on the SLC parameters of bcc Fe and also reported a significant impact. Moreover, this naturally has implications on the Heisenberg-like and DMI-like contributions to the magnetoelastic energy density, as described in Eq. \ref{eq:complete-magnetoelastic-expression} (Appendix \ref{sec:continuum-limit}).

The emerging interest in understanding the angular momentum flow between the spin and lattice subsystems \cite{Tauchert2022} calls for a closer look into magnon-number nonconserving processes -- strongly linked to spin-lattice terms of spin-orbit coupling origin. An important quantity, argued in Ref. \cite{Mankovsky2022} to be physically related to the torque exchanged between the spin and lattice subsystems by a magnon-phonon interconversion process \cite{Ruckriegel2020}, is $\mathcal{D}_{\mathbf{q}}^{\alpha\mu}$ ($\alpha=x,y$), obtained by Fourier transforming the real-space $\mathcal{D}_{ijk}^{\alpha\mu}$ parameters \footnote{In systems with inversion symmetry, the quantity $\mathcal{D}_{\mathbf{q}}^{\alpha\mu}$ is purely imaginary. Numerically, we verify that $\frac{\textnormal{Re}(\mathcal{D}_{\mathbf{q}}^{\alpha\mu})}{\textnormal{Im}(\mathcal{D}_{\mathbf{q}}^{\alpha\mu})}<1\%$.}:

\begin{equation}
\label{eq:dq-parameters}
\mathcal{D}_{\mathbf{q}}^{\alpha\mu}=\sum_{jk}\mathcal{D}_{ijk}^{\alpha\mu}e^{i\mathbf{q}\cdot(\mathbf{r}_j-\mathbf{r}_i)}e^{-i\mathbf{q}\cdot(\mathbf{r}_k-\mathbf{r}_i)}.
\end{equation}

To better understand that, one can begin with the atomistic DMI-like Hamiltonian $\mathcal{H}_{\textnormal{DMI-like}}=\frac{1}{S^2}\sum_{ijk\mu}\boldsymbol{\mathcal{D}}_{ijk}^{\mu}\cdot\left(\hat{\boldsymbol{\mathcal{S}}}_i\times\hat{\boldsymbol{\mathcal{S}}}_j\right)$, where the classical $\leftrightarrow$ quantum spin substitution $m_i^{\alpha}\rightarrow\hat{\mathcal{S}}_i^{\alpha}/S$ is considered (here, $\hat{\boldsymbol{\mathcal{S}}}_i$ is the spin vector represented in terms of its Hermitian Cartesian component operators, and $S=\left|\langle\hat{\boldsymbol{\mathcal{S}}}\rangle\right|$). By applying the standard linear Holstein-Primakoff transformations (LHPT) ($\hat{\mathcal{S}}_i^{+(-)}\sim\sqrt{2S}\hat{b}_i^{(\dagger)}$, and $\hat{\mathcal{S}_i^{z}}=S-\hat{b}_i^{\dagger}\hat{b}_i$), and with the help of ladder definitions $\hat{\mathcal{S}}_i^{\pm}=\left(\hat{\mathcal{S}}_i^x\pm i\hat{\mathcal{S}}_i^y\right)$ and $\mathcal{D}_{ijk}^{\pm,\mu}=\left(\mathcal{D}_{ijk}^{x,\mu}\pm i\mathcal{D}_{ijk}^{y,\mu}\right)$, it can be shown that, in momentum space, $\mathcal{H}_{\textnormal{DMI-like}}$ is converted to \cite{Mankovsky2022} 

\begin{equation}
\label{eq:hme-dmi}
\begin{split}
\mathcal{H}_{\textnormal{DMI-like}}=\frac{2i}{\sqrt{2S}}\sum_{\mathbf{q}\mu}\left(\mathcal{D}_{\mathbf{q}}^{-,\mu}\hat{b}_{\mathbf{q}}-\mathcal{D}_{-\mathbf{q}}^{+,\mu}\hat{b}_{-\mathbf{q}}^{\dagger}\right)u_{\mathbf{q}}^{\mu} -\\
\frac{2i}{S\sqrt{N}}\sum_{\mathbf{k}\mathbf{k}^{\prime}\mu}\mathcal{D}_{\mathbf{k},\mathbf{k}^{\prime}}^{z,\mu}\hat{b}_{\mathbf{k}}^{\dagger}\hat{b}_{\mathbf{k}^{\prime}}u_{(\mathbf{k}-\mathbf{k}^{\prime})}^{\mu},
\end{split}
\end{equation}

\noindent where $\hat{b}_{\mathbf{k}}$ ($\hat{b}_{\mathbf{k}}^{\dagger}$) denote the annihilation (creation) operator of magnon with wavelenght $\mathbf{k}$, $N$ is total number of atoms in the lattice, and  $\mathcal{D}_{\mathbf{q}}^{\alpha\mu}$ follows the definition in Eq. \ref{eq:dq-parameters}. The first term in Eq. \ref{eq:hme-dmi} corresponds to magnon-number non-conserving scattering processes, which are associated with the transfer of angular momentum from the magnetic order to the lattice. The second term represents magnon-number \textit{conserving} processes, which are responsible solely for energy transfer \cite{Ruckriegel2020,Mankovsky2022}. Therefore, the coefficients $\mathcal{D}_{\mathbf{q}}^{\alpha\mu}$ dictate the magnitude of the one-phonon/one-magnon couplings involved in these processes. 

In this sense, in Fig.  \ref{fig:spin-lattice-temperature}  we also present the imaginary part of these coefficients for the case of $\alpha=\hat{x}$. From the behavior of $\textnormal{Im}(\mathcal{D}_{\mathbf{q}}^{x\mu})$, one can readly notice two features: (\textit{i}) a strong dependence on the wavevector $\mathbf{q}$; and (\textit{ii}) the existence of different $\textnormal{Im}(\mathcal{D}_{\mathbf{q}}^{x\mu})$ profiles, which depend on the $\mu$ displacements and indicate the distinct magnon couplings with transversal and longitudinal phonon modes. The additional symmetry breaking induced by the noncollinearity modifies the original profile of the curves, specially for displacements in the $\mu=\hat{x}$ direction, towards which the spin is canting. As expected from the real-space results, while  $\textnormal{Im}(\mathcal{D}_{\mathbf{q}}^{x\mu})$ is hardly changed for Co and Ni, in Fe this quantity is strongly affected. This means that the coupling strength related to the hybridization between phonons and magnons is, in general, strongly enhanced in Fe because of the noncollinearity induced by finite spin temperatures. Such modifications have a direct impact on the generation of magnetoelastic waves \cite{Ruckriegel2020}, and the angular momentum transfer between spins and the lattice, being an effect to be accounted in more realistic atomistic dynamics simulations -- such as those pursued recently by Korniienko \textit{et al.} \cite{Korniienko2024}.

Another noteworthy quantity, also of spin-orbit coupling origin, is the \textit{nonlocal} contribution to the magnetic anisotropy that arises from an atomic displacement. Analogously to the equilibrium parameter discussed by Udvardi \textit{et al.} \cite{Udvardi2003}, we can define $\mathcal{K}_{ijk}^{\mu}=\frac{1}{2}\left(\Gamma_{ijk}^{xx\mu}-\Gamma_{ijk}^{yy\mu}\right)$ and its Fourier transform $\mathcal{K}_{\mathbf{k},\mathbf{k}^{\prime}}^{\mu}=\sum_{jk}\mathcal{K}_{ijk}^{\mu}e^{i\mathbf{k}\cdot(\mathbf{r}_j-\mathbf{r}_k)}e^{-i\mathbf{k}^{\prime}\cdot(\mathbf{r}_k-\mathbf{r}_i)}$. This quantity naturally originates from the consideration of the general term $\propto\sum_{ijk\mu}\Gamma_{ijk}^{xx\mu}m_i^{x}m_j^{x}u_k^{\mu}+\Gamma_{ijk}^{yy\mu}m_i^{y}m_j^{y}u_k^{\mu}$ that is part of the full spin-lattice Hamiltonian (Eq. \ref{eq:spin-lattice-hamiltonian-at}), and vanishes in the nonrelativistic case. An application of the LHPT leads to terms in which $\mathcal{K}_{\mathbf{k},\mathbf{k}^{\prime}}^{\mu(*)}$ are coupling strengths of number nonconserving magnon-phonon scattering processes involving two magnons and one phonon:

\begin{equation}
\label{eq:two-magnon-one-phonon}
\begin{split}
\mathcal{H}_{\textnormal{me-ani}}=\frac{1}{S\sqrt{N}}\sum_{\mathbf{k}\mathbf{k}^{\prime}\mu}(\mathcal{K}_{\mathbf{k},\mathbf{k}^{\prime}}^{\mu}\hat{b}_{\mathbf{k}}\hat{b}_{\mathbf{k}^{\prime}}u_{-(\mathbf{k}+\mathbf{k}^{\prime})}^{\mu}+\\\mathcal{K}_{\mathbf{k},\mathbf{k}^{\prime}}^{\mu *}\hat{b}_{\mathbf{k}}^{\dagger}\hat{b}_{\mathbf{k}^{\prime}}^{\dagger}u_{(\mathbf{k}+\mathbf{k}^{\prime})}^{\mu}).
\end{split}
\end{equation}

\begin{figure}[!htb]
\includegraphics[width=0.7\columnwidth]{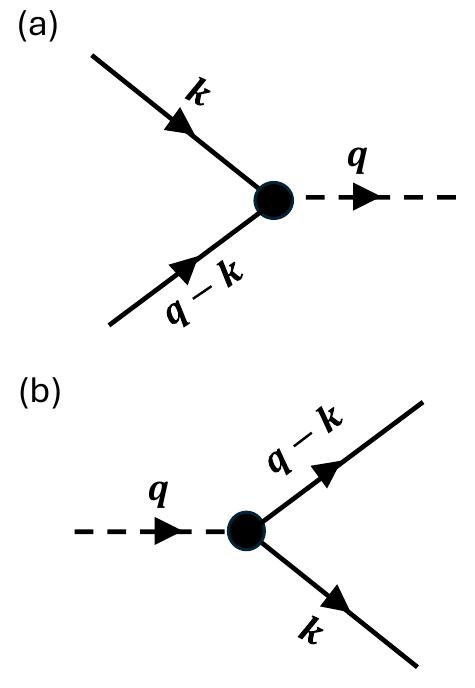}
\caption{Feynman diagrams illustrating number nonconserving magnon-phonon scattering processes involving two magnons (solid lines) and one phonon (dashed lines): (a) confluence; (b) splitting.}
\label{fig:confluence-splitting-process}
\end{figure}

Those are commonly referred to as magnon \textit{confluence} or \textit{splitting} processes \cite{Ruckriegel2020} (see Fig. \ref{fig:confluence-splitting-process} for a diagramatic representation), and also contribute to the transfer of angular momentum from the magnetic order to the lattice. Thus, the obtained $\textnormal{Im}(\mathcal{K}_{\mathbf{k},\mathbf{k}^{\prime}}^{\mu})$ curves can be seen in Appendix \ref{sec:nonlocal-anisotropy}, presenting a complex dependence on the pairs of momenta $(\mathbf{k},\mathbf{k}^{\prime})$. In accordance with Ref. \cite{Mankovsky2023}, this contribution is in general much smaller when compared to those coming from the DMI-like in all $3d$ ferromagnets. However, when noncollinearity is considered, the values are greatly enhanced, resulting in a non-negligible contribution from the nonlocal anisotropy to the (total) angular momentum transfer. In this aspect, we notice that in certain materials, such as CoF$_2$, such two magnon-one phonon interactions can be particularly strong even at very low temperatures \cite{Metzger2024}. 

\subsection{Effect of pressure: the case of bcc iron}
\label{sec:pressure-effects}

Controlling the magnon-phonon coupling by external stimuli (what we can readly define as magnon-phonon engineering) is desirable by the application point of view. On one side, recent experimental investigations have established that high pressure can induce strong magnon-phonon interactions, as it is the case of the bulk layered FePS$_3$ \cite{Pawbake2022}. Such strong magnon-phonon coupling can be of interest in many ways, such as to the formation of the so-called magnon-polarons \cite{Kamra2015,Streib2018}, to enable coherent information processing that combines the advantages of both spintronics and straintronics \cite{Li2021}, or even to the efficient manipulation of magnetic skyrmions via acoustic waves \cite{Yang2024}. On the other side, inducing a weaker magnon-phonon coupling diminishes the exchange-mediated magnetoelastic contribution to the damping of magnons (specially those of shorter wavelengths), directly impacting in their lifetimes \cite{Hellsvik2019,Streib2019}. From the results presented in the previous Section (\ref{sec:temperature}), and based on the peculiar competition between inter-orbital exchange interactions to which it is subjected \cite{Kvashnin2016}, bcc Fe stands out from the $3d$ ferromagnets as an interesting candidate for the analysis of the effect of pressure in the atomistic spin-lattice parameters.

At moderate temperatures ($T\lesssim 1000$ K) and low pressures, iron crystallizes in the $\alpha$-phase (bcc), transforming into the $\varepsilon$-phase (hcp) when the external pressure reaches $\sim 11$ GPa \cite{Anderson1997}. However, even beyond this thermodynamic stability limit, the $\alpha$-phase continue to be metastable; the bcc phase only becomes dynamically unstable at much higher pressures ($P\sim180$ GPa \cite{Vocadlo2003}), and the dynamical precursor effects of the Burgers mechanism \cite{Burgers1934}, associated with the bcc-hcp transition, are still absent \cite{Klotz2000}. Thus, it is reasonable to present theoretical results even for $P\gtrsim11$ GPa, specially because our aim here is to investigate the behavior of spin-lattice couplings under more extreme conditions in iron.

\begin{figure*}[!htb]
\includegraphics[width=1.5\columnwidth]{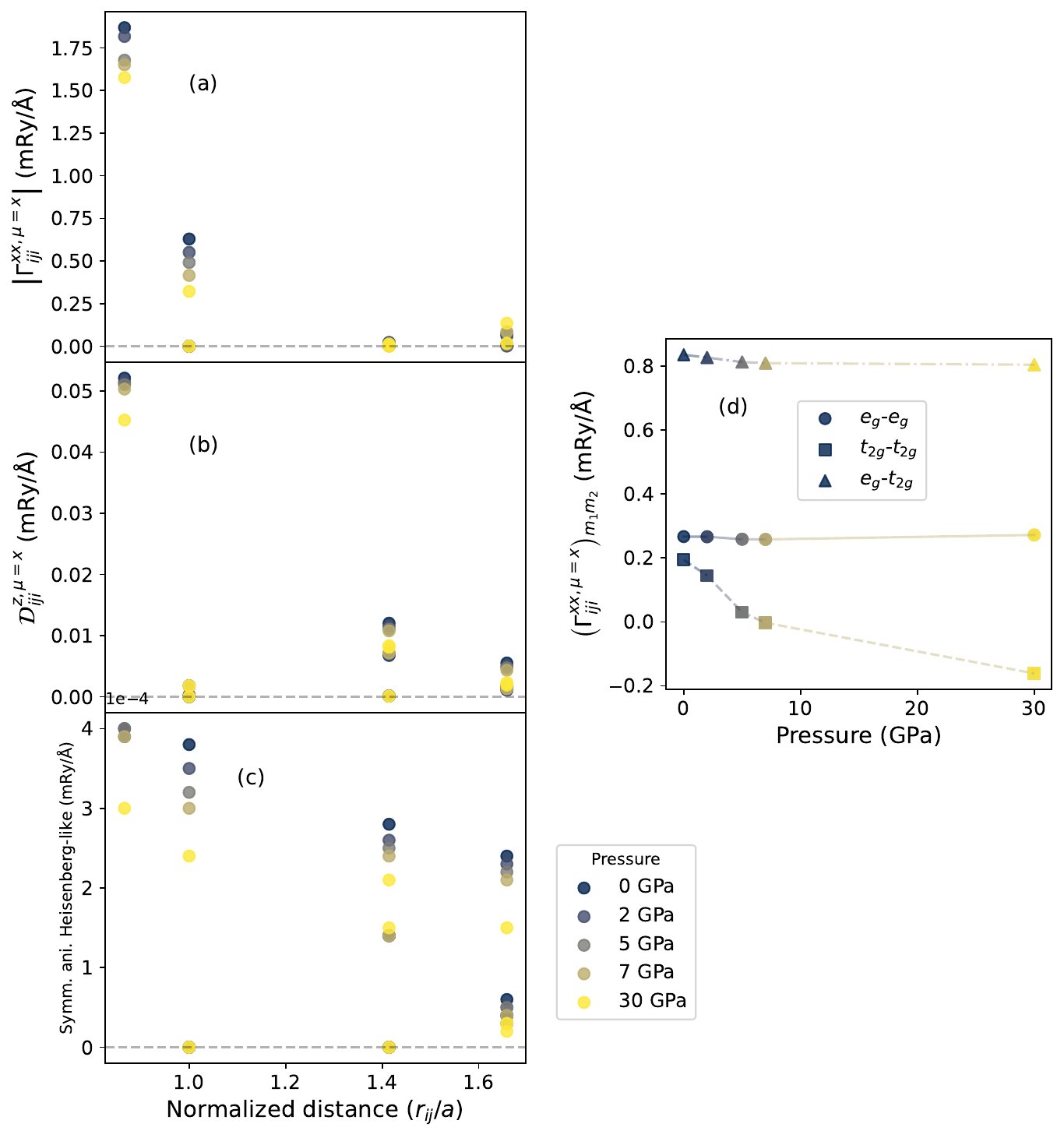}
\caption{Change in the spin-lattice coupling parameters of bcc Fe when an isotropic contraction (isodeformation in all three directions) is applied, as a function of the normalized interatomic distance $\frac{r_{ij}}{a}$: (a) diagonal SLC parameters for $k=i$; (b) DMI-like couplings (also for $k=i$); (c) Symmetric anisotropic Heisenberg-like interactions induced by the displacement of the $i$-th atom. (d) Orbital decomposed NN diagonal SLC parameter as a function of the applied pressure. All values expressed for a SQA in $\hat{z}$ and displacement in the $\mu=\hat{x}$ direction. Lines are guides for the eyes.}
\label{fig:pressure-effect}
\end{figure*}

When pressures of the order of $0\leq P\lesssim1.75$ GPa are applied to pure iron, an intriguing phenomenon occurs: the Curie temperature stays almost constant, differently from other ferromagnetic materials, such as nickel \cite{Leger1972}. At the core of this observation lies two competing factors: the simultaneous increase of the exchange interaction parameters and reduction of local magnetic moments \cite{Kormann2009}, expected as the ferromagnetic state tends to become more itinerant as the interatomic distances decrease. This dichotomy is also obtained theoretically here, and present even for higher values of $P$ (see Appendix \ref{sec:appendix-pressure}). However, spin-lattice coupling seems to follow the opposite trend in Fe: although a contraction induces larger spin-spin interactions, spin-lattice becomes weaker. Figure \ref{fig:pressure-effect} shows the results for the SLC parameters when a pressure $P\in[0,30]$ GPa corresponding to an isotropic contraction is applied to bcc Fe, together with the orbital decomposition of the nearest-neighbor diagonal (Heisenberg-like) interaction. We immediately see that the diagonal SLC parameters, in the $k=i$ case, diminish by $\sim17\%$, which is, in good approximation, followed by the corresponding off-diagonal terms ($\mathcal{D}_{iji}^{z\mu}$ and $\varGamma_{iji}^{z\mu}$). From the conversion discussed in Section \ref{sec:connection-measurements}, this results an absolute decrease of the $k=i$ contribution to the magnetoelastic parameters $B^{\prime}_{\parallel}$ and $B^{\prime}_{\perp}$. Since the contributions for $k=i$ are negative, particularly for the diagonal elements $B^{\prime}_{\parallel}$, this consequently increases their values: from 0 GPa to 30 GPa, $B^{\prime}_{\parallel}$ roughly doubles, while $B^{\prime}_{\perp}$ becomes less negative by $\sim20\%$. In turn, the calculation of $\textnormal{Im}(\mathcal{D}_{\mathbf{q}}^{\alpha\mu})$ shows a clear trend, as the curves decrease almost everywhere with increasing pressure in the whole Brillouin Zone, with a maximum decrease of $\sim20\%$ for $\mu=\hat{y}$ (data not shown). Such modified parameters by pressure (including now the DMI-like) influence the dispersion relations, affecting both the gap observed around the crossing points of magnon and phonon bands \cite{Ruckriegel2014}, induced by their interaction, and the attenuation of both spin \cite{Streib2019} and sound \cite{Lord1968} waves in the material.  

To inspect the origin of such effect, we again relied on the orbital decomposition of the SLC parameters, here analyzed for the nearest-neighbor Heisenberg-like interactions (Fig. \ref{fig:pressure-effect}(d)). Unlike the static situation, where the mixed and short-ranged $e_g$-$t_{2g}$ interactions are majorly responsible for the value of $\Gamma_{iji}^{\alpha\alpha\mu}$ (Section \ref{sec:analysis-3d-ferromagnets}), its change (for first neighbors) when an external pressure is applied has the primary contribution coming from the long-range, RKKY-type, $t_{2g}$-$t_{2g}$ interactions -- while $e_g$-$e_g$ and $e_g$-$t_{2g}$ spin-lattice couplings, mainly characterized by the double-exchange mechanism \cite{Kvashnin2016}, are much less sensible to such structural modifications.

\section{Conclusions and outlook}

 We have extensively investigated and compared the spin-lattice coupling parameters of the $3d$ elemental ferromagnets (Fe, Co, Ni) with \textit{ab-initio} accuracy. The calculations were performed within the proposed simplified embedded cluster approach, which shows excellent agreement with the (computationally more expensive) fully self-consistent scheme. From the point of view of the ground-state structures and magnetization, we demonstrate the existence of not only diagonal Heisenberg-like, but also symmetric and antisymmetric anisotropic (Dzyaloshinskii-Moriya-like) interactions, of spin-orbit coupling origin. When noncollinearity (roughly related to spin temperature) is taken into account, these parameters, especially in Fe, can undergo significant changes; this has serious implications for the generation of magnetoelastic waves and the transfer of angular momentum between the spin and lattice subsystems. Additionally, for Fe under isotropic contraction, there is a notable change in the exchange-mediated magnetoelastic constants, including a significant enhancement of $B_{\parallel}^{\prime}$. In the nearest-neighborhood, this is shown to be influenced by long-range RKKY-type interactions, and therefore driven by changes in Fermi surface topology due to atomic displacements. Finally, a conversion between the atomistic, quantum-mechanically derived, coupling parameters and the phenomenological magnetoelastic theory is presented and applied to the calculation of the $B_{\parallel}^{\prime}$ and $B_{\perp}^{\prime}$ constants, to which we find that the inclusion of nonlocal interactions ($k\neq i(j)$) is crucial.

 The implications of this investigation are diverse. First, the analysis presented here builds upon existing research and may serve as a benchmark for future experimental and theoretical studies in the field. Second, our findings expand the current knowledge of the spin-lattice-induced Dzyaloshinskii-Moriya interaction as the primary mechanism for angular momentum transfer between the magnetic system and the lattice, demonstrating that other actors can play a crucial role in the presence of noncollinearity. Third, while we show that the commonly employed (and computationally affordable) $J_{ij}(r_{ij})$ expansion significantly misrepresents the underlying physics of spin-lattice interactions, we explore the connection between atomistic coupling parameters and continuum magnetoelastic theory beyond magnetostriction. This approach can enable large-scale simulations grounded in \textit{ab-initio} results, incorporating both isotropic and anisotropic terms, offering a viable alternative route for this purpose.
\section{Acknowledgments}

I.P.M. thanks S. Streib for  early discussions. The work was financially supported by the Crafoord Foundation (Grant No. 20231063), the Knut and Alice Wallenberg Foundation (grant numbers 2018.0060, 2021.0246, and 2022.0108), and the Wallenberg Initiative Materials Science for Sustainability (WISE)
funded by the Knut and Alice Wallenberg Foundation. A.B. acknowledges support from eSSENCE and Carl Tryggers Foundation. Support from the Swedish Energy Agency (Energimyndigheten), the European Research Council (854843-FASTCORR), eSSENCE, and STandUP is acknowledged by O.E. M. Pankratova acknowledges support from the Olle Engkvist Foundation. A.B.K. acknowledges support from CNPq, FAPESPA, the INCT of Materials Informatics, and Spintronics and Advanced Magnetic Nanostructures. Financial support from Swedish Research Council (VR) (grant numbers 2016-05980, 2019-05304, 2019-03666, 2023-04239, and 2024-04986) is acknowledged by D.T., O.E.~and A.D. The computations/data handling were enabled by resources provided by the National Academic Infrastructure for Supercomputing in Sweden (NAISS), partially funded by the Swedish Research Council through grant agreement no. 2022-06725.


\appendix

\section{Realistic calculation of atomic displacements}
\label{sec:realistic-atomic-displacements}

\begin{figure}[!htb]
\includegraphics[width=0.5\textwidth]{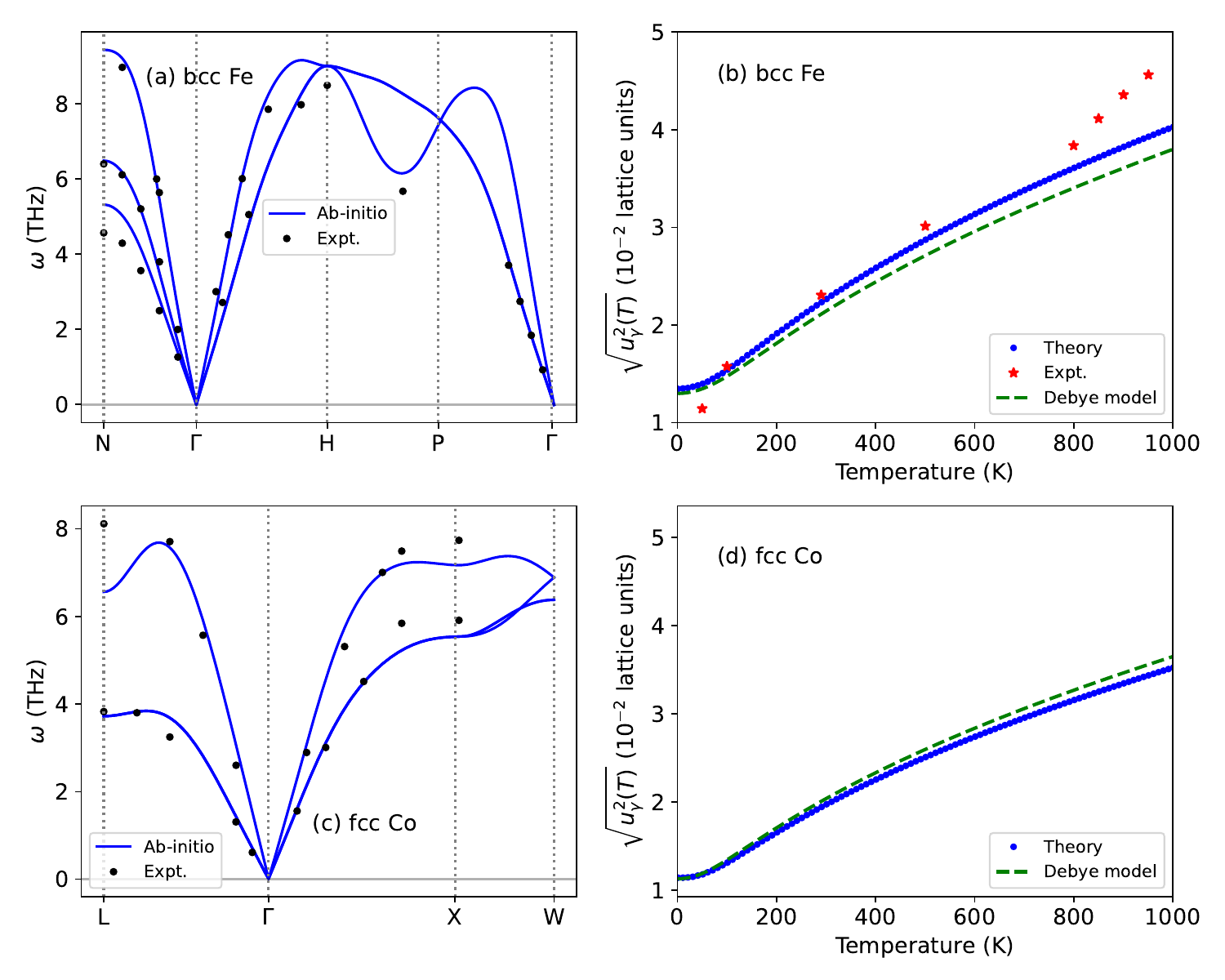}
\caption{(a,c) Computed ferromagnetic phonon dispersions of bcc Fe and fcc Co, and comparison with room-temperature experimental data (black dots) extracted from Ref. \cite{Klotz2000} (bcc Fe) and Ref. \cite{Shapiro1977} (fcc Co$_{0.92}$Fe$_{0.08}$); (b,d) Correspondent root-mean square (RMS) thermal displacements calculated within the HA (blue dots), together with the estimation by the Debye theory (Eq. \ref{eq:debye-model}, green lines), for the experimental values $\theta_D=470.2$ K (bcc Fe, from Ref. \cite{Hill2008}), and $\theta_D=385$ K (fcc Co, from Ref. \cite{Armentrout2015}). In (b) we also show the RMS displacements experimental data from Ref. \cite{Owen1967} (red stars), where the raw values were divided by $\sqrt{3}$ as we are plotting only one of the Cartesian components.}
\label{fig:phonon-displacements}
\end{figure}

The expectation value of thermal atomic displacements, $\langle u^2\rangle$, correspondent to a given lattice temperature can be theoretically obtained from several methods. In the present work, we use the thermal displacements within the harmonic approximation (HA), as implemented in PHONOPY \cite{Togo2015}, and which is dependent on the phonon frequencies calculated by means of the \textit{ab-initio} force constants. It is widely known that the HA is usually good for solid state systems at sufficiently low temperatures, from which we calculate the maximum displacement for determining the spin-lattice parameters. An example for bcc Fe and fcc Co is shown in Figure \ref{fig:phonon-displacements}, together with experimental data from Refs. \cite{Klotz2000,Schmalzl2006} and a comparison with the values estimated using Debye's theory: 

\begin{equation}
\label{eq:debye-model}
\langle u^2\rangle=\frac{1}{4}\frac{3h^2}{\pi^2mk_B\theta_D}\left(\frac{T}{\theta_D}\Phi(\theta_D/T)+\frac{1}{4}\right),
\end{equation}
 
 \noindent where $\theta_D$ is the Debye temperature (here taken as experimental results), $\Phi(x)=\int_{z=0}^{x}\frac{z}{e^{z}-1}dz$ is the Debye function, $m$ the atomic mass, $h$ the Plack constant, and $k_B$ represents the Boltzmann constant. which has been used before, and can also represent a rather good approximation for $\langle u^2\rangle$. Similar calculations and analysis were performed for fcc Ni, and are already reported in Ref. \cite{Pankratova2022}.

\section{Atomistic spin-lattice Hamiltonian in the continuum limit}
\label{sec:continuum-limit}

We start deriving the continuum limit of the atomistic spin-lattice Hamiltonian by considering an equivalent expression of Eq. \ref{eq:spin-lattice-hamiltonian-at} to account for the \textit{relative} atomic displacements $\left(u^{\mu}_k-u^{\mu}_i\right)$ \footnote{Note that if $u_i^{\mu}=0$, i.e., if the $i$-th atom is at its initial equilibrium position, we recover Eq. \ref{eq:spin-lattice-hamiltonian}.}, as already considered in Refs. \cite{Weissenhofer2023,Wojdel2013}:

\begin{equation}
\label{eq:spin-lattice-hamiltonian}
\mathcal{H}_{SL}=-\frac{1}{2}\sum_{ijk}\sum_{\alpha\beta\mu}\Gamma_{ijk}^{\alpha\beta\mu}m_i^{\alpha}m_j^{\beta}\left(u^{\mu}_k-u^{\mu}_i\right).
\end{equation}

Let us first consider the case $\alpha=\beta$, in which Eq. \ref{eq:spin-lattice-hamiltonian} can be rewritten as

\begin{equation}
\label{eq:spin-lattice-hamiltonian2}
\mathcal{H}_{SL}=-\frac{1}{2}\sum_{ijk}\sum_{\mu\alpha}\Gamma_{ijk}^{\alpha\alpha\mu}\mathbf{m}_i\cdot\mathbf{m}_j\left(u^{\mu}_k-u^{\mu}_i\right).
\end{equation}

In this case, the local moment $\mathbf{m}_{j}\equiv \mathbf{m}(\mathbf{r}_{j})$ can be expanded as $\mathbf{m}_{j}\sim\left[\mathbf{m}+(\mathbf{r}_{ij}\cdot\nabla)\mathbf{m}+\frac{1}{2}(\mathbf{r}_{ij}\cdot\nabla)^2\mathbf{m}\right]$ (for $\mathbf{r}_{ij}=\mathbf{r}_i-\mathbf{r}_j$), and the relative atomic displacements $\mathbf{u}_k-\mathbf{u}_i\sim(\mathbf{r}_{ik}\cdot\nabla)\mathbf{u}$ up to the first order approximation. Now for $\mathbf{m}_j$, the first term gives a constant contribution to the energy, as $\left|\mathbf{m}\right|^2=1$. The second term vanishes as one can show that $\mathbf{m}\cdot(\mathbf{r}_{ij}\cdot\nabla)\mathbf{m}=\sum_{\kappa\eta}r_{ij}^{\kappa}m^{\eta}\frac{\partial m^{\eta}}{\partial x_{\kappa}}=\frac{1}{2}\sum_{\kappa\eta}r_{ij}^{\kappa}\frac{\partial}{\partial x_{\kappa}}\left[(m^{\eta})^2\right]=\frac{1}{2}\sum_{\kappa}r_{ij}^{\kappa}\frac{\partial}{\partial x_{\kappa}}\left[\sum_{\eta}(m^{\eta})^2\right]=0$, for the same reason as the first term ($\left|\mathbf{m}\right|^2=1$). Thus, using the substitution $\sum_i\rightarrow\frac{1}{V}\int d\mathbf{r}$ and assuming that the boundary terms vanish in the integration by parts, Eq. \ref{eq:spin-lattice-hamiltonian2} can be rewritten as

\begin{equation}
\label{eq:hsl-continuum-alpha-alpha}
\mathcal{H}_{SL}\sim\frac{1}{4V}\int d\mathbf{r}\sum_{jk}\sum_{\mu\alpha\kappa\eta\zeta\tau}\Gamma_{ijk}^{\alpha\alpha\mu}r_{ij}^{\zeta}r_{ij}^{\tau}r_{ik}^{\kappa}\frac{\partial m^{\eta}}{\partial x_{\zeta}}\frac{\partial m^{\eta}}{\partial x_{\tau}}\frac{\partial u^{\mu}}{\partial x_{\kappa}},
\end{equation}

\noindent where $V$ is the volume of the unit cell (occupied by each atom $i$), and the tensor

\begin{equation}
\label{eq:tensor-a-prime}
A^{\prime}_{\alpha,\mu\kappa\eta\zeta\tau}=\frac{1}{4V}\sum_{jk}\Gamma_{ijk}^{\alpha\alpha\mu}r_{ij}^{\zeta}r_{ij}^{\tau}r_{ik}^{\kappa},
\end{equation}

\noindent can be defined. We immediately see that the parameters defined by Eq. \ref{eq:tensor-a-prime}, for large distances, are proportional to $\sim r^3$, which might make the series converge slowly (or even diverge) as a function of neighbors for itinerant-electron systems (as $\Gamma_{ijk}^{\alpha\alpha\mu}$ decays as $\sim\frac{1}{r^4}$, see Section \ref{sec:analysis-3d-ferromagnets}). This is essentially the same problem that appeared when trying to calculate the spin-wave stiffness constant from atomistic $J_{ij}$'s \cite{Pajda2001} -- for which we can employ an analogous solution:

\begin{equation}
\label{eq:damped-tensor-a-prime}
A^{\prime}_{\alpha,\mu\kappa\eta\zeta\tau}=\lim_{\Lambda\rightarrow 0}\frac{1}{4V}\sum_{jk}\Gamma_{ijk}^{\alpha\alpha\mu}r_{ij}^{\zeta}r_{ij}^{\tau}r_{ik}^{\kappa}e^{-\Lambda\bar{r}},
\end{equation}

\noindent for $\Lambda \in \mathbb{R}_{\geq 0}$. Here, we choose to write the attenuation factor dependent on the average of the distances $\bar{r}=\frac{1}{2}\left(\left|\mathbf{r}_{ij}\right|+\left|\mathbf{r}_{ik}\right|\right)$ due to the fact that $\Gamma_{ijk}^{\alpha\alpha\mu}$ interactions are thought to be influenced by the combined effect of the distances.

Now we aim to compare the expression of Eq. \ref{eq:hsl-continuum-alpha-alpha} to the classical magnetoelastic energy density expression \cite{Streib2019,Kaganov1959,Abrahams1952,Kittel1949} (see Section \ref{sec:connection-measurements}):

\begin{equation}
\label{eq:magnetoelastic-energy}
\varepsilon_{me}=\int d\mathbf{r}\sum_{\mu\kappa}\left[B_{\mu\kappa}m^{\mu}m^{\kappa}+B^{\prime}_{\mu\kappa}\frac{\partial\mathbf{m}}{\partial x_{\mu}}\cdot\frac{\partial\mathbf{m}}{\partial x_{\kappa}}\right]\epsilon_{\mu\kappa},
\end{equation}

\noindent in which the strain tensor elements $\epsilon_{\mu\kappa}=\frac{1}{2}\left(\frac{\partial u^{\mu}}{\partial x_{\kappa}}+\frac{\partial u^{\kappa}}{\partial x_{\mu}}\right)$ are standard in infinitesimal tensor theory and defined in terms of the atomic displacements. Note that this expression does not take into account the volume magnetostriction term, independent of the magnetization direction and related to the exchange interactions in the Néel model \cite{Chikazumi2009,Neel1954}. In the spin-lattice dynamics simulations, this effect should be included via Eq. \ref{eq:spin-lattice-hamiltonian-at}, as it is usually simulated through a $J_{ij}(r_{ij})$ dependence in the spin Hamiltonian \cite{Tranchida2018}. With that in mind, it is possible to split the tensor $\mathbf{A}^{\prime}$ into symmetric ($+$) and antisymmetric ($-$) parts such that it matches with Eq. \ref{eq:magnetoelastic-energy}:

\begin{equation}
(A^{\prime}_{\alpha,\mu\kappa\eta\zeta\tau})^{\pm}=\frac{1}{8V}\sum_{jk}\left(\Gamma_{ijk}^{\alpha\alpha\mu}r_{ij}^{\zeta}r_{ij}^{\tau}r_{ik}^{\kappa}\pm\Gamma_{ijk}^{\alpha\alpha\kappa}r_{ij}^{\zeta}r_{ij}^{\tau}r_{ik}^{\mu}\right),
\end{equation}

\noindent from which one can easily write

\begin{widetext}
\begin{equation}
\begin{split}
\label{eq:demonstration1}
\mathcal{H}_{SL}\sim\int d\mathbf{r}\sum_{\mu\alpha\kappa\eta\zeta\tau}\left[(A^{\prime}_{\alpha,\mu\kappa\eta\zeta\tau})^{+}+
(A^{\prime}_{\alpha,\mu\kappa\eta\zeta\tau})^{-}\right]\frac{\partial m^{\eta}}{\partial x_{\zeta}}\frac{\partial m^{\eta}}{\partial x_{\tau}}\frac{\partial u^{\mu}}{\partial x_{\kappa}}=\\
\int d\mathbf{r}\sum_{\mu\alpha\kappa\eta\zeta\tau}\frac{\partial m^{\eta}}{\partial x_{\zeta}}\frac{\partial m^{\eta}}{\partial x_{\tau}}
\left[\frac{1}{2}(A^{\prime}_{\alpha,\mu\kappa\eta\zeta\tau})^{+}\left(\frac{\partial u^{\mu}}{\partial x_{\kappa}}+\frac{\partial u^{\kappa}}{\partial x_{\mu}}\right)+\frac{1}{2}(A^{\prime}_{\alpha,\mu\kappa\eta\zeta\tau})^{-}\left(\frac{\partial u^{\mu}}{\partial x_{\kappa}}-\frac{\partial u^{\kappa}}{\partial x_{\mu}}\right)\right]=\\
\int d\mathbf{r}\sum_{\mu\alpha\kappa\eta\zeta\tau}\frac{\partial m^{\eta}}{\partial x_{\zeta}}\frac{\partial m^{\eta}}{\partial x_{\tau}}\left[(A^{\prime}_{\alpha,\mu\kappa\eta\zeta\tau})^{+}\epsilon_{\mu\kappa}+(A^{\prime}_{\alpha,\mu\kappa\eta\zeta\tau})^{-}\omega_{\mu\kappa}\right]
\end{split}
\end{equation}
\end{widetext}

Thus, the magnetoelastic constants $B_{\mu\kappa}^{\prime}$ due to the exchange interactions under lattice deformations can be written as

\begin{equation}
\label{eq:magnetoelastic-constant-b-prime}
B^{\prime}_{\mu\kappa}=\lim_{\Lambda\rightarrow0}\frac{1}{8V}\sum_{\alpha jk}\left(\Gamma_{ijk}^{\alpha\alpha\mu}r_{ij}^{\mu}r_{ij}^{\kappa}r_{ik}^{\kappa}+\Gamma_{ijk}^{\alpha\alpha\kappa}r_{ij}^{\kappa}r_{ij}^{\mu}r_{ik}^{\mu}\right)e^{-\Lambda\bar{r}},
\end{equation}

\noindent where $e^{-\Lambda\bar{r}}$ is the same \textit{ad hoc} attenuation factor introduced in Eq. \ref{eq:tensor-a-prime}. Note that this decomposition leads to a much richer physics than described by early magnetoelastic theories, including also naturally terms dependent on the infinitesimal rotation tensor $\omega_{\mu\kappa}=\frac{1}{2}\left(\frac{\partial u^{\mu}}{\partial x_{\kappa}}-\frac{\partial u^{\kappa}}{\partial x_{\mu}}\right)$, previously investigated by, e.g., Bonsall and Melcher \cite{Bonsall1976} and Bar'yakhtar \textit{et al.} \cite{Baryakhtar1985}. This term has been associated, for instance, with differences in the speeds of transverse sounds propagating along the easy axis and perpendicular axes of certain materials, such as the antiferromaget MnF$_2$ \cite{Melcher1970}. It is also interesting to see how they directly connect with the quantum-mechanically derived $\Gamma_{ijk}^{\alpha\alpha\mu}$ parameters.

For $i=j$ (i.e., for the on-site interaction, as previously defined, for instance, by Antropov \textit{et al.} \cite{Antropov1999} and Mankovsky \textit{et al.} \cite{Mankovsky2023}), then it is possible to write the magnetostriction terms correspondent to $B_{\mu\kappa}$ in Eq. \ref{eq:magnetoelastic-energy} as 

\begin{equation}
\label{eq:magnetocaloric-constant-b}
B_{\mu\kappa}=\lim_{\Lambda\rightarrow0}\frac{1}{4V}\sum_{k}\left(\Gamma_{iik}^{\mu\kappa\mu}r_{ik}^{\kappa}+\Gamma_{iik}^{\mu\kappa\kappa}r_{ik}^{\mu}\right)e^{-\Lambda r_{ik}},
\end{equation}

\noindent by using the same reasoning of separating in symmetric and antisymmetric parts the resulting tensor $A_{\alpha\beta,\mu\kappa}=\frac{1}{2V}\sum_{k}\Gamma_{iik}^{\alpha\beta\mu}r_{ik}^{\kappa}$ in this case. Thus, the $B_{\mu\kappa}$ elements, related to the local $\Gamma_{iik}^{\alpha\beta\mu}$ terms and as discussed previously \cite{Mankovsky2022}, describe effects induced by spin-orbit coupling when the local symmetry is broken by the displacement of the $k$-th atom. Within the Néel model \cite{Neel1954}, they are strongly related to the dipole term.   Interestingly, in the collinear-\textit{nonrelativistic} limit the local $J_{ii}$ terms can be shown to vanish when calculated from one-site and two-site rotation approaches within the total variation of the grand potential \cite{Szilva2023}. However, in a spin-orbit coupling environment where also atoms are allowed to move from their equilibrium positions, this symmetric cancellation is not expected to hold anymore.

Up to now, only the case $\alpha=\beta$ has been considered for $i\neq j$. However, if we allow also for $\alpha\neq\beta$, then both Dzyaloshinskii-Moriya (when spin-orbit coupling is enabled) and symmetric anisotropic Heisenberg interactions \cite{Borisov2023}, will intuitively lead to contributions to the magnetoelastic energy density. Although the former has been only very recently considered by Weißenhofer \textit{et al.} \cite{Weissenhofer2023}, the latter -- to the best of our knowledge -- has been never addressed (even though it naturally leads to an analogous expression to the DMI-like contribution); in particular, the symmetric anisotropic Heisenberg couplings can be relevant in the context of complex atomic structures, such as the kagome $A$V$_3$Sb$_5$ (where $A$ is either Cs, Rb, K) compounds \cite{Karmakar2023,Hasan2023}. In order to exoress those contributions, one may expand $\Gamma_{ijk}^{\alpha\beta\mu}$, for $\alpha\neq\beta$, as $\Gamma_{ijk}^{\alpha\beta\mu}=\sum_{\gamma}\varepsilon_{\alpha\beta\gamma}(\mathcal{D}_{ijk}^{\gamma\mu}+\varGamma_{ijk}^{\gamma\mu})$, where $\varepsilon_{\alpha\beta\gamma}$ is the Levi-Civita symbol, and $\varGamma_{ijk}^{\alpha\beta\mu}$ represents the symmetric anisotropic exchange coupling induced by the (relative) displacement of the $k$-th atom in the $\mu$ direction. In turn, $\mathbf{m}_{j}$ may be expanded to its first order in $\mathbf{m}$ ($\mathbf{m}_{j}\sim\mathbf{m}+(\mathbf{r}_{ij}\cdot\nabla)\mathbf{m}$), as the first order terms do not vanish anymore as in the case of $\alpha=\beta$. 
 
Therefore, in general, Eq. \ref{eq:spin-lattice-hamiltonian} can be translated into the following (\textit{extended}) energy density expression using again the same symmetric/antisymmetric separation of the resulting tensors:

\begin{widetext}
\begin{equation}
\label{eq:complete-magnetoelastic-expression}
\begin{split}
\varepsilon_{SL}\sim\underbrace{\int d\mathbf{r}\sum_{\alpha\beta\mu\kappa}m^{\alpha}m^{\beta}\left[(A_{\alpha\beta,\mu\kappa})^{+}\epsilon_{\mu\kappa}+(A_{\alpha\beta,\mu\kappa})^{-}\omega_{\mu\kappa}\right]}_{\textnormal{contribution from the on-site induced anisotropies}\,\Gamma_{iik}^{\alpha\beta\mu}}+\\
\underbrace{\int d\mathbf{r}\sum_{\mu\alpha\kappa\eta\zeta\tau}\frac{\partial m^{\eta}}{\partial x_{\zeta}}\frac{\partial m^{\eta}}{\partial x_{\tau}}\left[(A^{\prime}_{\alpha,\mu\kappa\eta\zeta\tau})^{+}\epsilon_{\mu\kappa}+(A^{\prime}_{\alpha,\mu\kappa\eta\zeta\tau})^{-}\omega_{\mu\kappa}\right]}_{\textnormal{contribution from the Heisenberg-like interactions}\,\Gamma_{ijk}^{\alpha\beta\mu}}+\underbrace{\int d\mathbf{r}\sum_{\mu\alpha\kappa\eta\zeta\tau}\varepsilon_{\alpha\beta\gamma}m^{\alpha}\frac{\partial}{\partial x_{\zeta}}m^{\beta}\left[(C_{\gamma\zeta,\mu\kappa})^{+}\epsilon_{\mu\kappa}+(C_{\gamma\zeta,\mu\kappa})^{-}\omega_{\mu\kappa}\right]}_{\textnormal{contribution from the induced DMI-like}\,\mathcal{D}_{ijk}^{\gamma\mu}}+\\
\underbrace{\int d\mathbf{r}\sum_{\mu\alpha\kappa\eta\zeta\tau}\varepsilon_{\alpha\beta\gamma}m^{\alpha}\frac{\partial}{\partial x_{\zeta}}m^{\beta}\left[(C_{\gamma\zeta,\mu\kappa}^{\prime})^{+}\epsilon_{\mu\kappa}+(C_{\gamma\zeta,\mu\kappa}^{\prime})^{-}\omega_{\mu\kappa}\right]}_{\textnormal{contribution from the induced symmetric-anisotropic Heisenberg-like}\,\varGamma_{ijk}^{\gamma\mu}},
\end{split}
\end{equation}
\end{widetext}

\noindent where the elements of the tensors $\mathbf{C}$ and $\mathbf{C}^{\prime}$, introduced in Eq. \ref{eq:complete-magnetoelastic-expression}, can be written as

\begin{equation}
\label{eq:elements-b-and-b-prime}
\begin{split}
(C_{\gamma\zeta,\mu\kappa})^{\pm}=\lim_{\Lambda\rightarrow0}\frac{1}{4V}\sum_{jk}\left(\mathcal{D}_{ijk}^{\gamma\mu}r_{ij}^{\zeta}r_{ik}^{\kappa}\pm\mathcal{D}_{ijk}^{\gamma\kappa}r_{ij}^{\zeta}r_{ik}^{\mu}\right)e^{-\Lambda\bar{r}} \\
(C_{\gamma\zeta,\mu\kappa}^{\prime})^{\pm}=\lim_{\Lambda\rightarrow0}\frac{1}{4V}\sum_{jk}\left(\varGamma_{ijk}^{\gamma\mu}r_{ij}^{\zeta}r_{ik}^{\kappa}\pm\varGamma_{ijk}^{\gamma\kappa}r_{ij}^{\zeta}r_{ik}^{\mu}\right)e^{-\Lambda\bar{r}}.
\end{split}
\end{equation}

Obviously, for centrosymmetric materials, the DM-like contribution to the magnetoelastic energy density present a \textit{dynamic} character (i.e., approximately vanishes for very low lattice temperatures). This contribution has been experimentally observed in, e.g., surface acoustic waves propagating in ultrathin CoFeB/Pt bilayers \cite{Kuss2020}. In Ref. \cite{Kuss2020}, the authors argue about a nonreciprocity effect provoked by DMI to be interesting for the realization of acoustic diodes. The terms in Eq. \ref{eq:complete-magnetoelastic-expression} can be further expanded by considering higher-order spin-lattice interactions (as, e.g., recently suggested in Ref. \cite{Mankovsky2023}).

\section{Calculation of $\bar{\theta}$ as a function of the spin temperature}
\label{sec:theta-bar-appendix}

The calculations were performed using the Uppsala Atomistic Spin Dynamics (UppASD) \cite{Skubic2008} package. The spin configuration, initially set as ferromagnetic (single-domain) inside a simulation cell of $30\times30\times30$ spins with periodic boundary conditions, was initially thermalized using the Monte Carlo method to each $T_s$. The spin Hamiltonian was set to consider both the equilibrium spin-spin exchange interactions, included up to a distance of $5a$ (fcc Ni, fcc Co) or $7a$ (bcc Fe), as well as the MAE. In such cubic systems, the energy $E$ as a function of the magnetization direction is well represented as $E=K_0+K_1(\alpha_1^2\alpha_2^2+\alpha_2^2\alpha_3^2+\alpha_3^2\alpha_1^2)$, where $\alpha_i$ ($i=\{1,2,3\}$) are the direction cosines. In our simulations, the constants $K_1$ were taken from experimental data: $K_1=3.9$ $\mu$eV/atom for bcc Fe (from Ref. \cite{fritsche1987relativistic}, with easy axis $[100]$), $K_1=8.1$ $\mu$eV/atom for fcc Ni (from Ref. \cite{AUBERT1977295}, with easy axis $[111]$), and $K_1=5.9$ $\mu$eV/atom for fcc Co (from Ref. \cite{PhysRevB.53.R10548}, with easy axis $[1\bar{1}1]$). The measurement phase (stochastic spin dynamics) to reach a representative average for the polar angle $\theta$ was simulated with $N=7\times10^{5}$ time steps of $dt=10^{-16}$ s, in an high damping regime ($\alpha=0.7$).

\begin{figure}[!htb]
\includegraphics[width=0.5\textwidth]{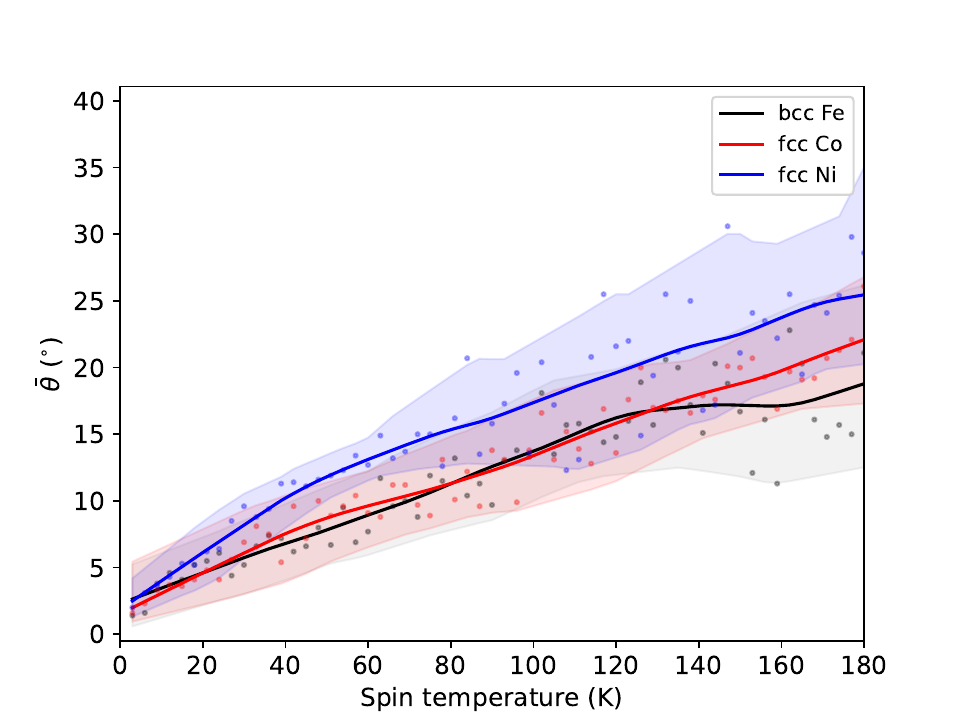}
\caption{Obtained $\bar{\theta}$ as a function of the spin temperature, $T_s$, for the $3d$ ferromagnets considered here. The points symbolize the raw data obtained directly from ASD simulations, while the full lines represent the smoothed function by the LOWESS algorithm \cite{Cleveland1979} with the parameter $f=0.25$.}
\label{fig:theta-temperature}
\end{figure}

Figure \ref{fig:theta-temperature} shows the obtained $\bar{\theta}$ values as a function of $T_s$ for bcc Fe, fcc Ni and fcc Co. To capture the essential shape of the $\bar{\theta}(T_s)$ curves and reduce the spread character induced by statistical noise and finite-time/finite-size measurements, we computed the smoothed function using a local weighted regression (LOWESS) algorithm \cite{Cleveland1979}. As we see, and as to be expected, the spread becomes larger upon increasing $T_s$, which restricts the region of confidence to low temperatures ($T_s\lesssim100$ K).

\section{Equation of state of iron and magnetic properties under pressure}
\label{sec:appendix-pressure}

\begin{figure*}[!htb]
\includegraphics[width=\textwidth]{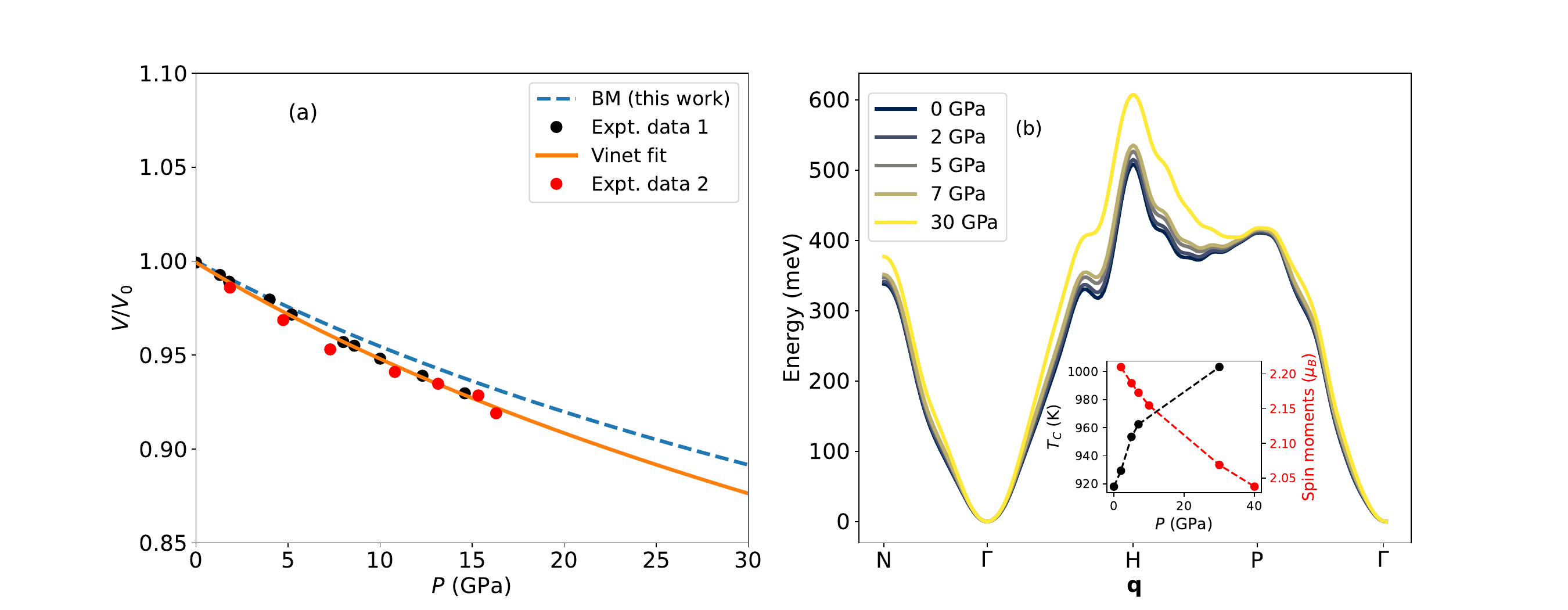}
\caption{(a) Equation of state for bcc Fe: calculated fully \textit{ab-initio} (dotted blue lines) and parameterized by a third-order Birch-Murnaghan function, and fitted to experimental data extracted from Ref. \cite{Dewaele2006} (black dots), parameterized by a Vinet function (full orange line). Red dots show another experimental set extracted from Ref. \cite{Jephcoat1986}. (b) adiabatic spin-wave spectra for different pressures applied to Fe. \textit{Inset:} change in the Curie temperature (black dots) obtained via RPA and the spin magnetic moment in $\mu_B$/atom (red dots) as a function of the applied pressure.}
\label{fig:pressure-properties}
\end{figure*}

The volume of bcc Fe ($\alpha$-Fe) under pressure was obtained by fitting the computed data points to the well-known third-order Birch-Murnaghan (BM) equation of state (EOS). The parameters obtained, namely the bulk modulus $K_0$, $K_0^{\prime}$ as its first pressure derivative, and the equilibrium volume $V_0$ are shown in Table \ref{tab:eos-iron}, and compared with other available data in the literature. Figure \ref{fig:pressure-properties}(a) presents the pressure-volume relation obtained -- used to calculate the lattice parameter for each compressed situation -- and compared to experimental data. Moreover, the evolution of the spin-wave spectra, spin magnetic moments and Curie temperature with applied pressure are displayed in Fig. \ref{fig:pressure-properties}(b).

From Fig. \ref{fig:pressure-properties}(a), we see that the fully \textit{ab-initio} results obtained here, although reasonably close, tend to overestimate the relative volume $\frac{V}{V_0}$. This is discussed in detail in Ref. \cite{Liu2024}. In summary, it can be seen as a deficiency of the PBE functional, and remedied with a better treatment of the electronic correlations. However, in view of the results presented in Section \ref{sec:pressure-effects}, as the measured $\frac{V}{V_0}$ relation is actually smaller than the calculated, we expect even larger effects.

Interestingly, early experimental results on Fe reported an almost constant pressure dependence of the Curie temperature $\left(\frac{\partial T_C(P)}{\partial P}\sim0\right)$ for $P\lesssim1.75$ GPa, subsequently investigated by first-principles calculations \cite{Kormann2009,Moran2003}; above that pressure, when iron meets the $\alpha$-$\gamma$ boundary, a measurable variation $\left|\frac{\partial T_C(P)}{\partial P}\right|>0$ can be seen -- from which we obtained $\gamma_m$ in Section \ref{sec:connection-measurements}. The mechanism associated with this almost constant $T_C$ as a function of $P$ is the competing increase of exchange interactions (which is reflected on the enhancement of the spin wave dispersions, as can be seen in Fig. \ref{fig:pressure-properties}(b)), and decrease of the local magnetic moments. Here we reasonably reproduce such constancy of $T_C$ pressures up to $P=2$ GPa (see Fig. \ref{fig:pressure-properties}(b), \textit{Inset}), with a slope of $\frac{\partial T_C(P)}{\partial P}\sim5.3$ K/GPa, in excellent agreement with Ref. \cite{Kormann2009}. The calculation of each $T_C$ was performed within the random-phase approximation (RPA), providing more accurate predictions than the mean-field results.

\begin{table}[!htb]
\caption{Equation of state parameters obtained by fitting of both experimental contraction data or computed binding energy curves of $\alpha$-Fe.}
\begin{center}
\label{tab:eos-iron}
\begin{tabular}{c c c c c}
\hline 
\hline
&&& \\[-2.5mm] 
& $V_0$ (\AA$^3$)  & $K_0$ (GPa)  & $K_0^{\prime}$ & EOS \\ 
&&& \\[-3mm] \hline
&&& \\[-3mm] 	    
PW-GGA (this work) & 11.81 & 188 & 5.7 & BM \\
&&& \\[-3mm] 
LAPW-GGA \cite{Stixrude1994}  & 11.15	& 189  & 4.9 & BM  \\ 
&&& \\[-3mm] 
LMTO-GGA \cite{Sha2006}  & 11.39 	& 178 &  4.7  & Vinet \\
&&& \\[-3mm] 
Expt. \cite{Sha2006}  & 11.78 	& 172 &  5.0  & Vinet \\
&&& \\[-3mm] 
Expt. \cite{Dewaele2006}  & 11.76 	& 167 &  4.8  & Vinet \\
\\\hline\hline
    \end{tabular}
 \end{center}	
\end{table}

\section{Nonlocal contribution to the magnetic anisotropy}
\label{sec:nonlocal-anisotropy}

\begin{figure*}[htb!]
    \centering
    \vspace{+1em}
    \includegraphics[width=0.32\textwidth]{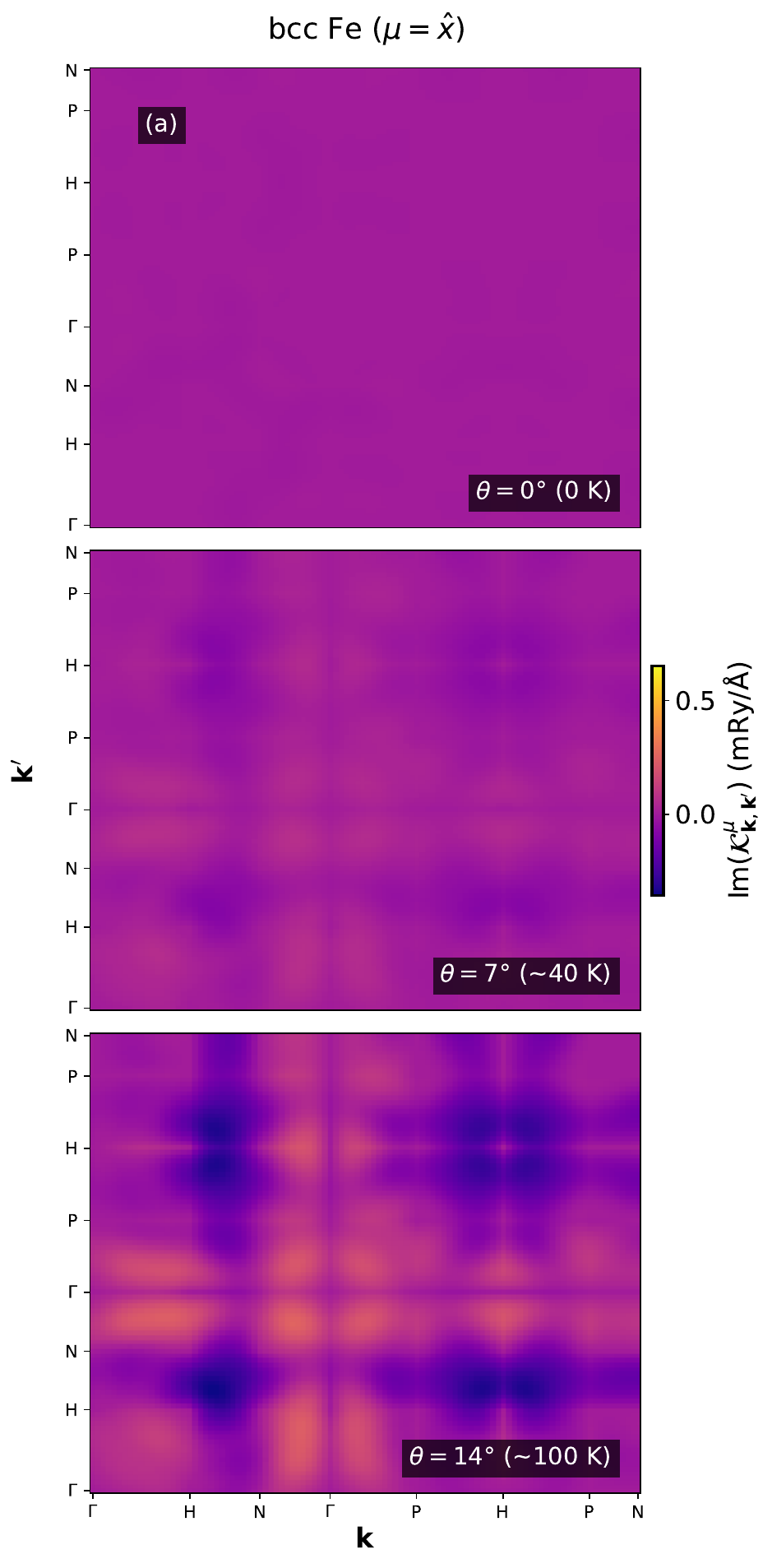}\hspace{-0.1cm}
    \includegraphics[width=0.32\textwidth]{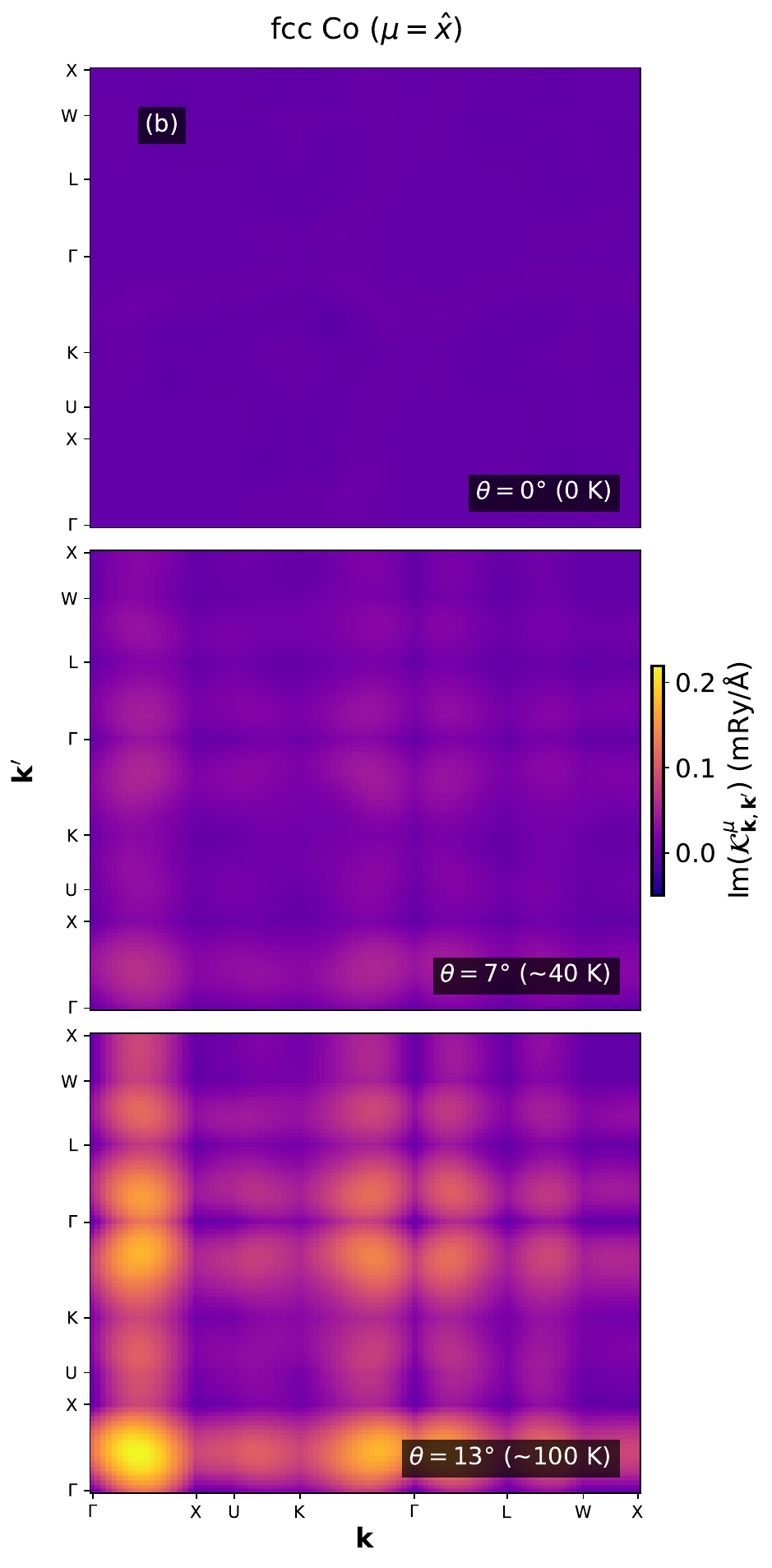}\hspace{-0.1cm}
    \includegraphics[width=0.335\textwidth]{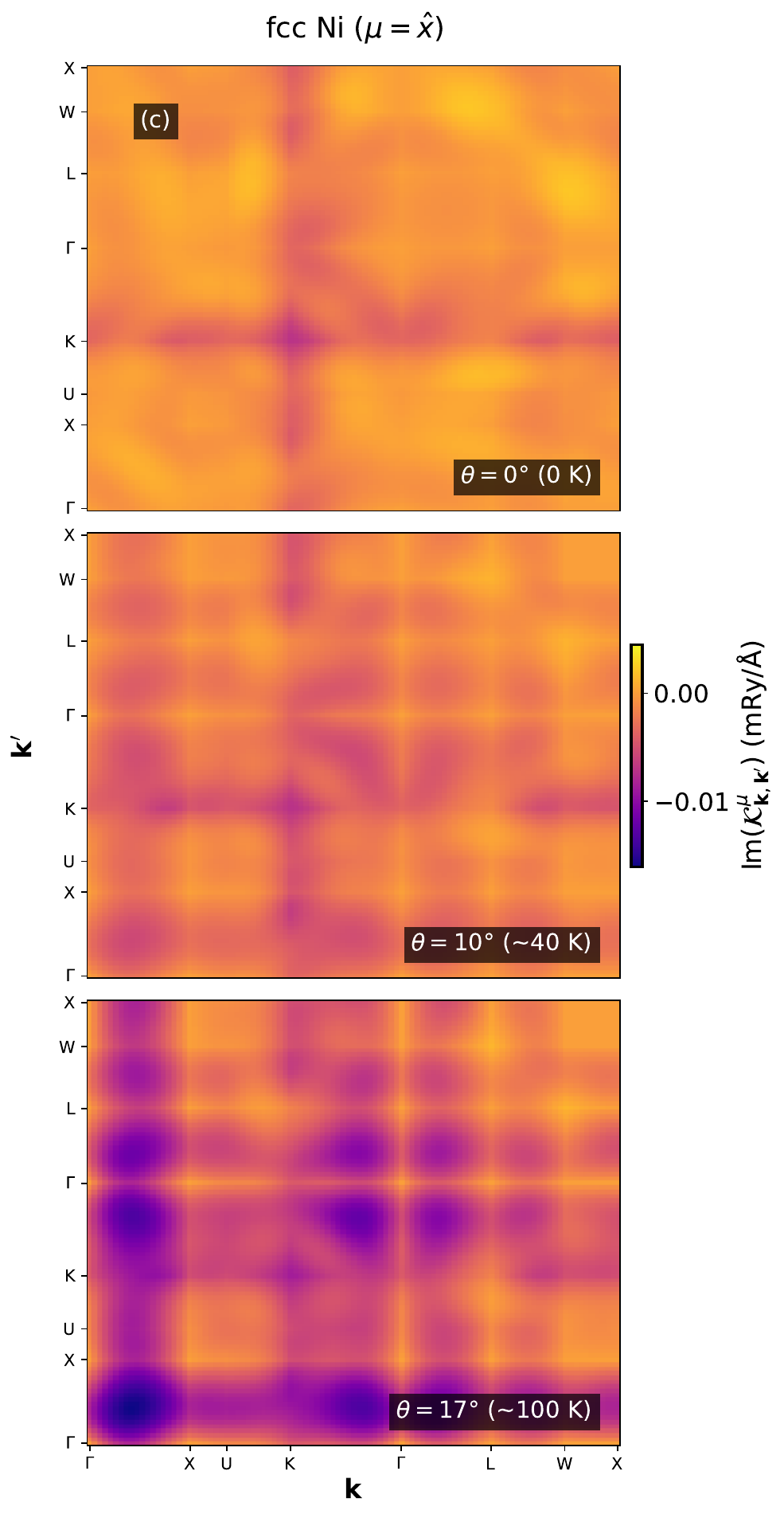}
    \caption{Imaginary part of the magnon-phonon coupling coefficient $\mathcal{K}_{\mathbf{k},\mathbf{k}^{\prime}}^{\mu}$, related to the real-space nonlocal anisotropy-like term $\frac{1}{2}\left(\Gamma_{ijk}^{xx\mu}-\Gamma_{ijk}^{yy\mu}\right)$ (see text). This anisotropy arises as a consequence of atomic displacements in each $\mu$ direction, and here is shown for $\mu=\hat{x}$ as a function of the polar angle $\bar{\theta}$ of the reference spin: (a) bcc Fe; (b) fcc Co; (c) fcc Ni.}
    \label{fig:gamma-temperature}
\end{figure*}

When an atom is displaced from its equilibrium position, the temporary and local breaking of inversion symmetry induces contributions to the magnetic anisotropy, driven by the existence of spin-orbit coupling. These contributions can be local as well as nonlocal in nature. The former has been extensively explored in the literature since the beginning of the phenomenological considerations of the magnetoelastic coupling \cite{Kittel1949}. In fact, it can be regarded as the leading term, and, as a consequence, more commonly considered in microscopic theories (see, e.g., Ref. \cite{Thingstad2019}). However, as pointed out by Udvardi \textit{et al.} \cite{Udvardi2003}, the mapping of the  energy of an itinerant electron system onto a spin Hamiltonian, in the relativistic case, naturally offers a contribution to the anisotropy energy emerging from \textit{nonlocal} interactions. In the static case, although not frequently of the leading order, this exchange anisotropy can be significant, depending on the lattice geometry or the presence of surfaces or interfaces \cite{Mankovsky2011}. In the context of spin-lattice interactions, the reasoning is analogous, but the presence of noncollinearity can generate strong modifications in the induced exchange anisotropy by atomic displacements -- thereby constituting a non-negligible contribution to the total spin-lattice anisotropy energy. In Figure \ref{fig:gamma-temperature} we demonstrate those effects in bcc Fe, fcc Co and fcc Ni, discussed in Section \ref{sec:temperature}.

\bigskip

\bibliographystyle{apsrev4-2}
\bibliography{apssamp}

\end{document}